\documentclass[preprint,12pt,authoryear]{elsarticle}

\usepackage{amssymb}

\usepackage{graphicx}
\usepackage{multirow}
\usepackage{siunitx}
\usepackage{booktabs}

\usepackage{xcolor}

\usepackage{amsmath}

\usepackage[hyphens]{url}
\usepackage{hyperref}
\usepackage[hyphenbreaks]{breakurl}
\usepackage{breakurl}

\journal{arXiV}

\begin{document}

\begin{frontmatter}



\title{Heart Failure Prediction using Modal Decomposition and Masked Autoencoders for Scarce Echocardiography Databases}


\author[inst1]{Andrés Bell-Navas\corref{cor1}}
\ead{a.bell@upm.es}

\author[inst3,inst4]{María Villalba-Orero}
\ead{mvorero@ucm.es}

\author[inst3]{Enrique Lara-Pezzi}
\ead{elara@cnic.es}

\author[inst1,inst5]{Jesús Garicano-Mena}
\ead{jesus.garicano.mena@upm.es}

\author[inst1,inst5]{Soledad {Le Clainche}}
\ead{soledad.leclainche@upm.es}

\cortext[cor1]{Corresponding author}

\affiliation[inst1]{organization={ETSI Aeronáutica y del Espacio, Universidad Politécnica de Madrid}, 
            addressline={Pl. del Cardenal Cisneros, 3}, 
            city={Madrid},
            postcode={28040}, 
            country={Spain}}

\affiliation[inst3]{organization={Centro Nacional de Investigaciones Cardiovasculares (CNIC)},
            addressline={C. de Melchor Fernández Almagro, 3}, 
            city={Madrid},
            postcode={28029}, 
            country={Spain}}

\affiliation[inst4]{organization={Departamento de Medicina y Cirugía Animal, Facultad de Veterinaria - Universidad Complutense de Madrid},
            addressline={Av. Puerta de Hierro}, 
            city={Madrid},
            postcode={28040}, 
            country={Spain}}
            
\affiliation[inst5]{organization={Center for Computational Simulation (CCS)},
            city={Boadilla del Monte},
            postcode={28660}, 
            country={Spain}}

\begin{abstract}
Heart diseases remain the leading cause of mortality worldwide, implying approximately 18 million deaths according to the WHO. In particular, heart failures (HF) press the healthcare industry to develop systems for their early, rapid, and effective prediction. This work presents an automatic system based on a novel framework which combines Modal Decomposition and Masked Autoencoders (MAE) to extend the application from heart disease classification to the more challenging and specific task of heart failure time prediction, not previously addressed to the best of authors' knowledge. This system comprises two stages. The first one transforms the data from a database of echocardiography video sequences into a large collection of annotated images compatible with the training phase of machine learning-based frameworks and deep learning-based ones. This stage includes the use of the Higher Order Dynamic Mode Decomposition (HODMD) algorithm for both data augmentation and feature extraction. The second stage builds and trains a Vision Transformer (ViT). MAEs based on a combined scheme of self-supervised (SSL) and supervised learning, so far barely explored in the literature about heart failure prediction, are adopted to effectively train the ViT from scratch, even with scarce databases. The designed neural network analyses in real-time images from echocardiography sequences to estimate the time of happening a heart failure. This approach demonstrates to improve prediction accuracy from scarce databases and to be superior to several established ViT and Convolutional Neural Network (CNN) architectures. The source code will be incorporated into the next version release of the ModelFLOWs-app software (\url{https://github.com/modelflows/ModelFLOWs-app}).
\end{abstract}


\end{frontmatter}


\section{Introduction}
\label{sec:intro}


Heart diseases and, in general, cardiovascular diseases (CVDs), remain the predominant cause of human mortality worldwide (\citet{Arooj2022deep}). A report conducted by the World Health Organization (WHO) shows that CVDs were responsible for nearly 18 million deaths in 2019, representing around a third of all global fatalities (\citet{who1999cvds}). Among these deaths, 85 \% were due to heart attacks and strokes. In addition, heart failures (HF) in particular affect around 24 million people worldwide with a median survival of five years (\citet{liu2023generalized}, \citet{valsaraj2023development}). This has become a great economic and social burden which increases over time due to the aging of populations, requiring demanding resources and high costs in the healthcare system (\citet{valsaraj2023development}). Therefore, early, rapid, and accurate identification and risk assessment of heart failures are of critical importance. Moreover, predicting the time of heart failure is essential for successful timely intervention and cost-effective treatments and, therefore, to improve the level of life. For heart failure prediction, echocardiography imaging is very widely used, specifically transthoracic echocardiography (TTE). This is because it contains much information about the structure and operation of the heart and, therefore, about the heart state and disease progression. In addition, it is a non-invasive sophisticated ultrasound method widely available, less costly, rapid, and non-ionizing, boosting low-resource settings and portability. However, the interpretation of echocardiography imaging, which must be performed by specialized clinicians, becomes challenging and subject to inter-observer variability, due to the quality and characteristics of the imaging (e.g., poor contrast, noise).

In recent years, Artificial Intelligence (AI) has shown significant potential to enhance healthcare, leading to automatic, faster, more accurate, and cost-effective diagnoses, minimizing human errors, and supporting clinical decision-making. Considering also the characteristics of echocardiography imaging and this global health crisis, the development of deep learning algorithms applied to echocardiography imaging has become of great interest for heart failure prediction. For example, the work in \citet{valsaraj2023development} adopts the ResNet architecture for a spatio-temporal Convolutional Neural Network (CNN) to predict HF mortality in 1, 3, or 5 years, obtaining accuracies of 81 \%, 75 \%, and 73 \%, respectively, in two private datasets. This work is based on the probability of defunction of patients in each of these timelines with respect to the acquisition date of the echocardiography video. The work considers the following subgroups: healthy, at risk of heart failure (HF), HF with reduced ejection fraction (HFrEF) and HF with preserved ejection fraction (HFpEF). The proposed model demonstrates to be superior to CatBoost gradient boosting based on echo measurements in external validation, i.e., in an independent dataset captured in another country, and so with other characteristics. Specifically, the areas under the receiver-operating curve (AUROC) obtained for the 1-, 3-, and 5-year mortality are 82 \%, 82 \%, and 78 \%, respectively, which are better than the 78\%, 73\%, and 75\% AUROCs obtained with CatBoost. Similarly, \citet{akerman2023automated} utilized a 3D CNN model for the diagnosis of HFpEF in TTE data from \num[group-separator={~}]{6823} patients, achieving an AUC of 91 \%. In \citet{zhang2018fully}, the proposed CNN automatically measures cardiac structure and function to compute LVEF, which has shown a median absolute difference of 6 \% with respect to manual tracings in more than \num[group-separator={~}]{14000} echocardiograms. In \citet{behnami2018automatic}, a dual-stream model was proposed for TTE videos from Apical two-chamber (A2C) and four-chamber (A4C) views. The architecture includes shared Recurrent Neural Network layers (RNN) and view-specific feature extraction blocks. This model directly estimates the left ventricle ejection fraction (LVEF) without previous LV segmentation or identification of crucial cardiac frames. Finally, a threshold of 40 \% applied on the estimated LVEF value determines either a high risk of heart failure (if below the threshold), or a low risk (otherwise). The work in \citet{liu2023generalized} proposes \textit{r2plus1d-Pan}, a deep spatio-temporal convolutional model compatible with both static and dynamic ultrasound images for the first time, for the diagnosis of HFrEF (i.e., again, EF below 40 \%). When trained with these two types of images, the model outperformed most human experts, comprising 15 registered ultrasonographers and cardiologists with different working years in three databases: EchoNet-Dynamic, Cardiac Acquisitions for Multi-structure (CAMUS) dataset, and a local one from the National Cardiovascular Center of China. However, the model results too large, supposing 57.2 GB of file storing.

Some works based on other deep learning techniques which estimate the LVEF have also been proposed. For instance, the work in \citet{muhtaseb2022echocotr} combines the advantages of 3D CNNs and Vision Transformers (ViTs) by proposing EchoCoTr, obtaining a Mean Absolute Error of 3.95 in the EchoNet-Dynamic dataset. This result is better than the one obtained in \citet{reynaud2021ultrasound} (5.95). Specifically, this work adopts a Transformer architecture based on a residual autoencoder network and on a modified version of the BERT (Bidirectional Encoder Representations from Transformers), acting as a spatio-temporal feature extractor. In parallel, in \citet{fazry2022hierarchical}, hierarchical ViTs are proposed, with a Mean Absolute Error of 5.59 in the LVEF estimate, without previous LV segmentation.

Among the previously presented works, CNNs are more frequently used for heart failure prediction (\citet{petmezas2024recent}), and this is based on determining their grade of risk (e.g., via the EF, as in \citet{behnami2018automatic}) or estimating the probability of mortality on a standard timeline (as in \citet{valsaraj2023development}). In addition, these works do not make adaptations to deal with the great challenge of scarcity of annotated samples, common in the medicine field.

On the other hand, our earlier work (\citet{bell2025automatic}) considers the more general, common application of AI to automated heart disease classification from echocardiography imaging. It introduces a framework which combines Modal Decomposition techniques, including a data-driven method, the Higher Order Dynamic Mode Decomposition (HODMD) (\citet{le2017higher}), and ViTs for classification of different heart states. That approach has demonstrated strong performance, becoming superior to several pretrained CNNs. However, again, like the previously presented works, that one does not address the clinically relevant question of specifically \emph{when} heart failures can happen.

Therefore, this work takes a significant step forward, extending our previous study, by formulating the heart failure prediction problem as a time-to-event prediction task, i.e., as a regression problem. In particular, this paper addresses the more specific task of quantitatively estimating the concrete age of patients in which heart failures will happen. For this reason, this task is more challenging than other heart disease recognition-based tasks. In addition, the complexity of cardiac dynamics makes this prediction of heart failures hard. This is because there could be subtle temporal and physiological changes which could correlate with disease progression, and thus with the event of heart failure. Therefore, these changes become crucial to capture. Our previous work addresses the classification between different heart states, but does not consider the gravity of the disease, meaning that the previously mentioned changes are not as relevant as in this work. The heart failure time prediction task is different and clinically more accurate and important than the usual and more general tasks of classifying between a set of heart states, or between levels of risk of heart failure (i.e., low, high) in a standard timeline. To the best of the authors' knowledge, this more challenging and specific task has not been addressed in the related literature, which implies direct clinical relevance in patient risk stratification. The contributions are summarized as follows:

\begin{itemize}
\item This paper introduces an extended data-driven solution, a novel framework for automatic heart failure time prediction in real-time which combines Modal Decomposition techniques, including the HODMD algorithm, and Masked Autoencoders (MAE) (\citet{he2022masked}).

\item A novel procedure which uses different sources of echocardiograms with heart failure prognoses, provided by human experts, to create a larger annotated database than the one from the previous study (\citet{bell2025automatic}). This procedure overcomes the limitations of that study in creating the database, more effectively addressing the heart failure prediction task in the usual scenario of scarcity of number of samples. This also broadens the application range of the proposed framework from general heart disease classification to the more challenging task of heart failure time prediction.

\item A joint self-supervised (SSL) and supervised learning scheme adopted by the MAE, barely explored to the best of the authors’ knowledge in the related literature on heart failure prediction in echocardiography images. This improves representation learning under scarce-data conditions, further addressing the difficulty of collecting large high-quality databases in the medicine field for a proper training, especially in ViTs (\citet{Lee2021vision}) in a more efficient and effective way than the modified ViT from the previous study (\citet{bell2025automatic}). 

\item An extensive evaluation using a created echocardiography database which compares several established CNNs and ViTs. This demonstrates the superiority of the proposed approach with a lower model complexity, which, in turn, enables efficient deployment on portable systems.

\end{itemize}

This work also largely extends our preliminary studies carried out in \citet{groun2025eigenhearts}, \citet{Bell2023HODMD}, and \citet{ESAO2024} by proposing the new database creation procedure, the deep learning architecture, and providing extensive experimental results on heart failure prediction performance. The source code used in this work is available in the next version release of the ModelFLOWs-app software (\citet{HETHERINGTON2024109217}). 

The remainder of the paper is organized as follows: Section \ref{sec:system} presents the details of the proposed heart failure prediction system. Section \ref{sec:database} introduces the database generated for the evaluation of the proposed system. Section \ref{sec:results} describes the experiments performed, summarizes the results obtained with the proposed system, and shows a comparison with other algorithms. Finally, conclusions are drawn in Section \ref{sec:conclusions}.

\section{Heart Failure Prediction System}
\label{sec:system}

This section describes the heart failure prediction system, which addresses the more challenging and specific task of predicting the time of happening a heart failure. To the best of the authors' knowledge, this task has not previously been addressed in the related literature. For this purpose, this framework analyses echocardiography images in real-time by combining Modal Decomposition techniques and MAEs. This is an extension of our previous work (\citet{bell2025automatic}), because it introduces several changes in the methodology to overcome the limitations encountered and to more effectively address the difficulty in the medicine field to elaborate a varied high-quality database to properly train deep learning algorithms. In addition, the proposed framework broadens the application range from the more common and widely studied tasks of classification between heart states or grades of risk of heart failure to the quantification of the time to heart failure. The information provided is more clinically relevant for decision-making processes than in heart disease classification (addressed in our previous work), as it helps to directly guide treatment planning. The output of this system is the prediction of the time of happening a heart failure from the given input sequence of echocardiography images.

This framework includes a novel procedure (described in Subsection \ref{sec:cardiac_data_creation}) to create a large annotated database from echocardiography video sequences; these sequences show hearts under different conditions and pathology progressions, and have different acquisition characteristics. In this way, the designed procedure mitigates potential discrepancies due to different acquisition devices used, and so in the imaging characteristics. This alleviates the need to collect large databases typically required by ViTs (\citet{Lee2021vision}), due to being very expensive and labor-intensive. Each of these sequences also includes an annotation of the time in which a heart failure happens, provided by health professionals.

\begin{figure*}[!ht]

\centering
\includegraphics[width=8 cm]{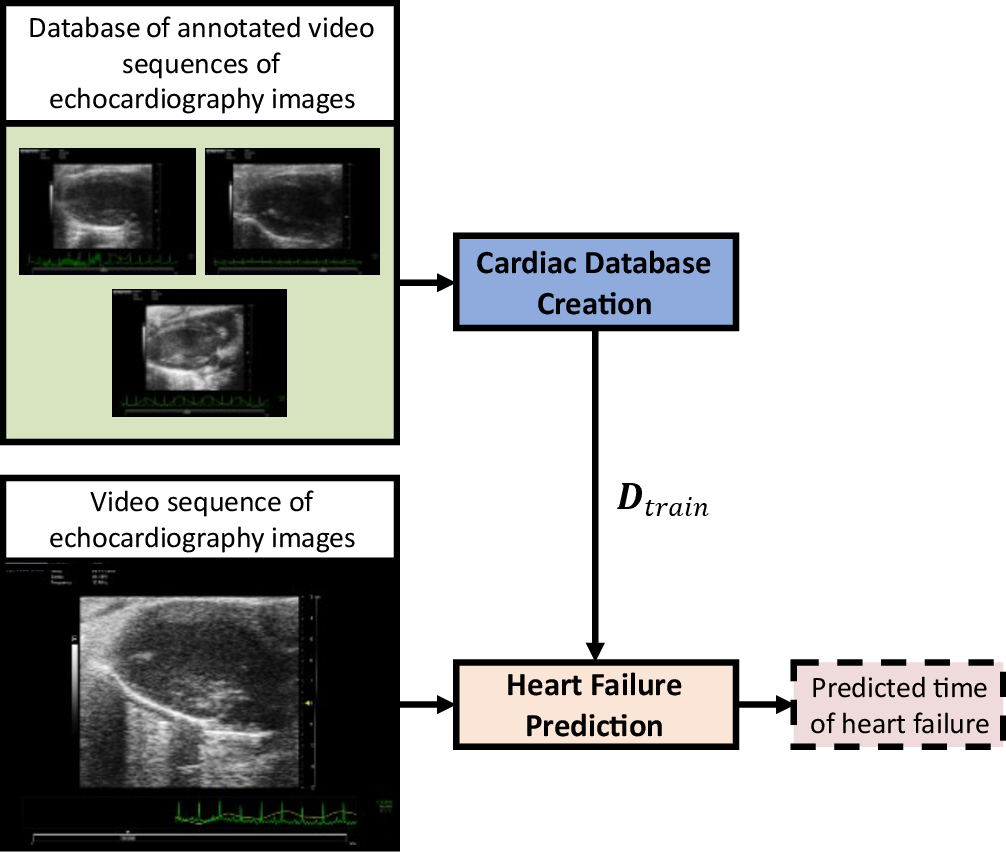}
\caption{Overall structure of the proposed heart failure prediction system, including representations of echocardiography images with non-heart regions (i.e., the electrocardiogram, medical information, and the black background).}
\label{fig:sys}
\end{figure*}

Fig. \ref{fig:sys} shows the overall structure of the proposed heart failure prediction system, comprising two main stages: Cardiac Database Creation, and Heart Failure Prediction. These are detailed in the following subsections.

\subsection{Cardiac Database Creation}
\label{sec:cardiac_data_creation}

This stage aims at creating a large annotated database from different sources of echocardiography images. The input is a set of sequences of echocardiography images, each one with annotations made by health professionals about the heart state, and also about the time of happening a heart failure. The output is an enlarged database of processed samples containing enriched low-noise representations of cardiac dynamics, which can be used to properly train machine learning and deep learning algorithms. The designed creation procedure here described overcomes the limitations encountered in our previous work (\citet{bell2025automatic}), because it generates more enriched samples; in contrast, our previous work could only leverage a subset of the generated samples for training. Therefore, the creation procedure here proposed addresses more effectively the prevalent difficulty in the medicine field to collect large varied high-quality databases for a proper training. This is because very hard specialized work with specific knowledge, which entails high costs, is required. This stage also includes a homogenization process to enable the use of the different sources of video sequences of echocardiography images, and improve the accuracy of standardized neural network architectures in the prediction of heart failures in the usual scenario of scarce samples. In addition, the proposed procedure broadens the application range of the created machine learning-compatible database to address the more challenging and specific task of predicting the time of heart failure, not previously addressed in the related literature to the best of the authors' knowledge. Moreover, the database creation procedure does not depend on the annotations included in the sequences of echocardiography images. This means that, with the proper annotations, the same created database can be leveraged to address different tasks in this context of heart disease recognition (e.g., heart disease classification, estimation of the LVEF), providing extended diagnostic information of a patient.

\begin{figure*}[!ht]

\centering
\includegraphics[width=14 cm]{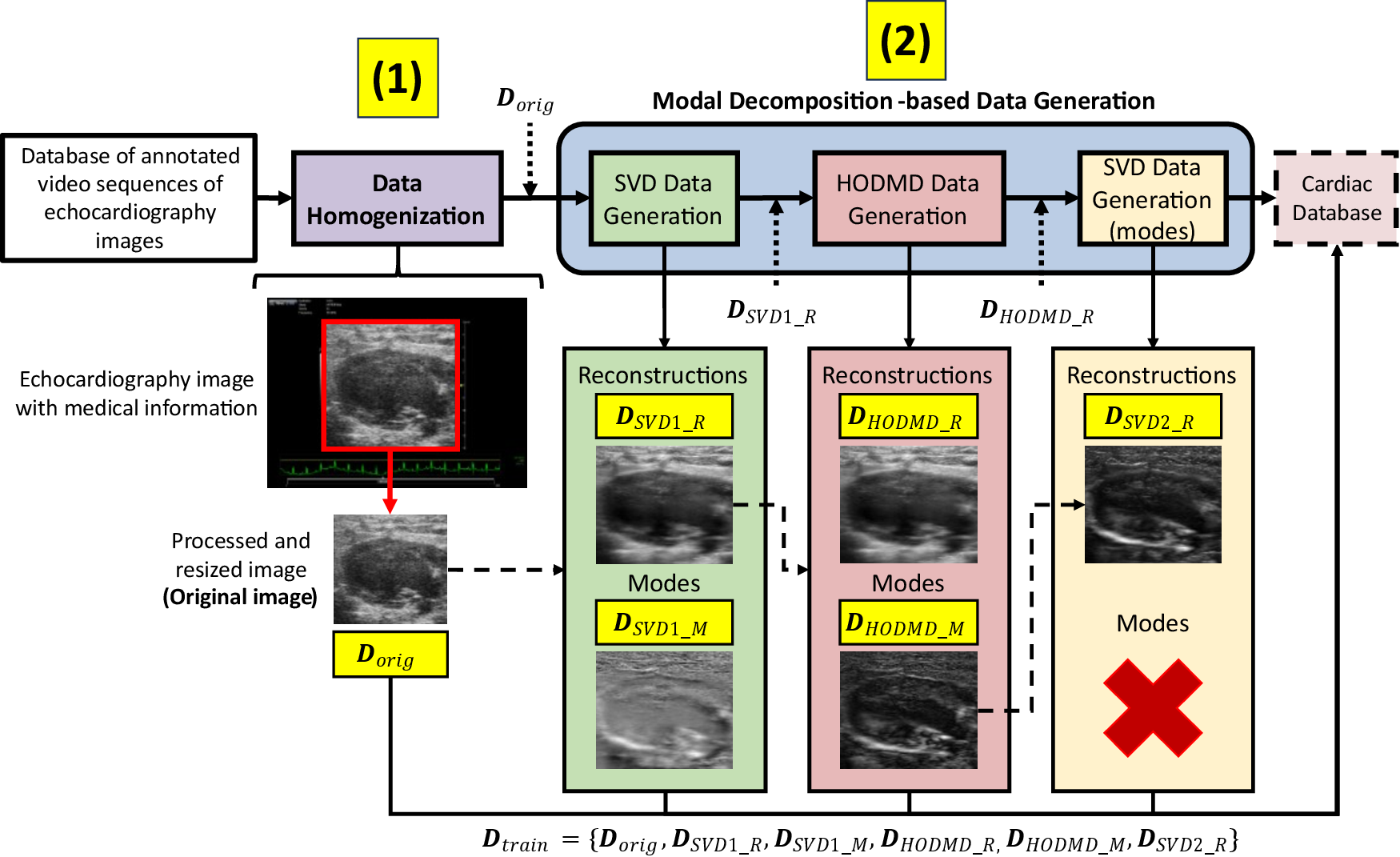}
\caption{Block diagram of the Cardiac Database Creation stage, composed of two phases: (1) Data Homogenization, and (2) Modal Decomposition-based Data Generation.}
\label{fig:cardiac_data_creation}
\end{figure*}

Fig.~\ref{fig:cardiac_data_creation} depicts the block diagram of the Cardiac Database Creation stage, which can be further divided into two phases: (1) Data Homogenization, and (2) Modal Decomposition-based Data Generation.

The first phase, Data Homogenization, extracts the region representing the heart from each image of the input sequence. The entire area outside this region is discarded (the one containing medical information and the black background). This is accomplished with edge detection algorithms in charge of locating the boundaries of the area corresponding to the echocardiography image itself (i.e., the region of interest, ROI, containing the heart).

The second and last phase, Modal Decomposition-based Data Generation, generates novel samples with less noise and containing features of the cardiac dynamics. The input comprises homogenized data (called 'original images', denoted as $\boldsymbol{D}_{orig}$) of the video sequences of echocardiography images. This generation is performed by sequentially applying the Singular Value Decomposition (SVD) (\citet{sirovich1987turbulence}) and the HODMD algorithms (\citet{le2017higher}, \citet{Groun2022higher}) to each homogenized sequence and taking the output of each algorithm, namely, reconstructions and modes. As Modal Decomposition techniques are employed for data augmentation, this phase acts as a machine learning method based on data physics. This is also complemented by conventional data augmentation techniques which apply geometric transformations (see Subsection \ref{sec:heart_fail_pred}). However, these conventional techniques only generate small variations of the actual samples and are not based on data physics. As they do not extract discriminative features which capture cardiac dynamics and the evolution of the heart state, they barely improve heart failure time prediction performance.  
The steps of the Modal Decomposition-based Data Generation phase are explained next. First, the SVD algorithm is applied to the homogenized sequences (i.e., $\boldsymbol{D}_{orig}$). This generates a set of modes ($\boldsymbol{D}_{SVD1\_M}$) and reconstructions of the original echocardiography images ($\boldsymbol{D}_{SVD1\_R}$). Next, the HODMD algorithm processes these reconstructions obtained with the previous SVD algorithm, i.e., $\boldsymbol{D}_{SVD1\_R}$. This generates modes and reconstructions, denoted as $\boldsymbol{D}_{HODMD\_M}$ and $\boldsymbol{D}_{HODMD\_R}$, respectively. For more details about the HODMD algorithm and its application in this work, see \S~\ref{sec:hodmd}. Finally, the SVD algorithm is again used, but on the HODMD modes ($\boldsymbol{D}_{HODMD\_R}$), also obtaining modes and reconstructions. However, only the reconstructions of the HODMD modes (as filtered versions) are considered ($\boldsymbol{D}_{SVD2\_R}$), while the SVD modes of these HODMD modes are discarded. Note that the HODMD algorithm can only be applied on sequences with a number of \textit{heartbeats} enough and, therefore, the following SVD algorithm. This is because the HODMD algorithm needs enough information about the cardiac function from a sequence to properly extract characteristic patterns and obtain fair reconstructions. For this reason, a threshold is imposed, which is the minimum number of snapshots a sequence must have to apply the HODMD algorithm.

In summary, reconstructions and modes are obtained from the previous steps. On the one hand, the reconstructions, as temporally consistent approximations of the original echocardiography images, have less noise and better represent the cardiac function and evolution. For this reason, these are considered for the database creation. Moreover, original echocardiography imaging is typically afflicted by noise, which hampers their interpretation. However, one of the main roles of the SVD and HODMD algorithms, as feature extractors, is to precisely filter the noise, better characterizing the heart state and evolution, and eventually improving the heart failure prediction performance. This means that the SVD and HODMD algorithms can also be seen as echocardiography video enhancement techniques, implying an important contribution to the relatively little research on techniques which enhance dynamic ultrasound video sequences (\citet{liu2025algorithms}). Moreover, these methods are usually based on supervised learning, rather than on data physics, unlike the SVD and HODMD algorithms. On the other hand, the HODMD modes (and their associated reconstructions obtained with the second use of the SVD algorithm) highlight relevant physical patterns which contain temporal information, representing the dynamics of the data and the cardiac behavior. These features can characterize heart diseases (\citet{Groun2022higher}), and -as we shall see- can also predict future heart failures.

The output of this stage is an enlarged annotated cardiac database, whose acquisition characteristics are homogenized to be adjusted to the target sensor used in the proposed heart failure prediction system. This forms the training database $\boldsymbol{D}_{train}$ for a deep neural network, the core of the Heart Failure Prediction stage described next, in Subsection \ref{sec:heart_fail_pred}. Table~\ref{tab:training_cases} summarizes the different combinations of the data generated in this stage (that is, $\boldsymbol{D}_{SVD1\_M}$, $\boldsymbol{D}_{SVD1\_R}$, $\boldsymbol{D}_{HODMD\_M}$, $\boldsymbol{D}_{HODMD\_R}$, and $\boldsymbol{D}_{SVD2\_R}$) which are taken to form $\boldsymbol{D}_{train}$. This table will be presented and described in more detail in Section \ref{sec:database}; we shall see that the combinations with precisely HODMD modes and reconstructions of the echocardiography images lead to the best results, because of having more samples and also more discriminative features derived from Modal Decomposition.

\subsubsection{Higher Order Dynamic Mode Decomposition}
\label{sec:hodmd}

The Higher Order Dynamic Mode Decomposition algorithm (HODMD) is a data-driven method (\citet{le2017higher}) which decomposes a time series into a set of modes, each with associated temporal dynamics, i.e., frequencies, growth rates, and amplitudes. This algorithm has been widely used in fluid dynamics and for different industrial applications (\citet{Groun2022higher}, \citet{VEGA202129}). This work focuses in particular on the application of the HODMD algorithm in the medicine field for data augmentation and feature extraction, supporting the training of the proposed deep neural network for heart failure time prediction. Specifically, the role of the HODMD algorithm in this work is to extract clinically meaningful patterns and capture temporal and physiological characteristics of the disease progression from video sequences of echocardiography images. This broadens the application range of the HODMD algorithm in the medicine field from heart disease classification (\citet{bell2025automatic}) to the more challenging and specific task of heart failure time prediction. From that decomposition, on the one hand, Dynamic Mode Decomposition (DMD) modes are obtained, where the most important ones contain temporally enriched representations of cardiac dynamics. These most relevant modes, according to \citet{Groun2022higher}, are typically those with the highest amplitudes (i.e., $a_{m}$ in Eq. \ref{eq:mode_exp}, represented next). Conversely, the modes with smaller amplitudes tend to be associated with non-relevant features, so with the high noise inherent to the echocardiography imaging, and thus with high frequencies. On the other hand, reconstructions of the original echocardiography images are obtained from the DMD modes.

\begin{figure*}[!ht]

\centering
\includegraphics[width=14 cm]{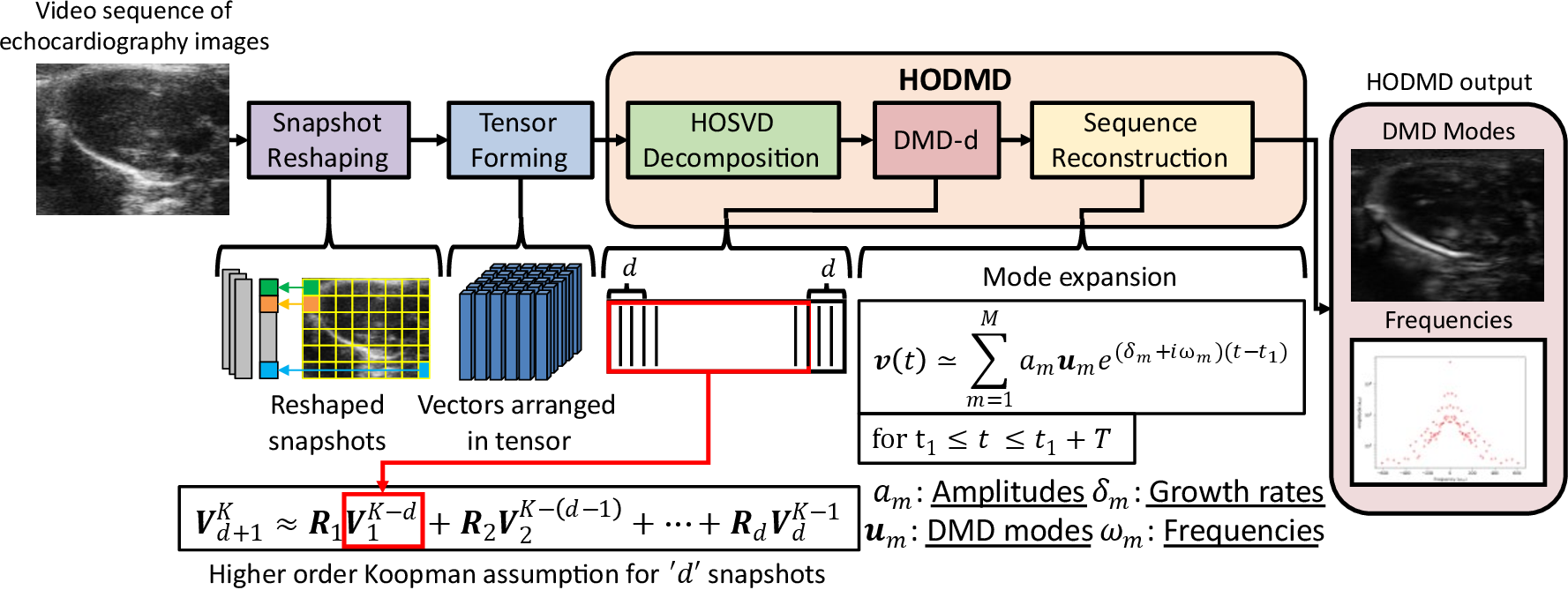}
\caption{Block diagram of the HODMD algorithm applied on a video sequence of echocardiography images.}
\label{fig:hodmd}
\end{figure*}

If only these most representative modes are considered and the noisy ones filtered out, the reconstructions will contain less noise and more discriminative features than the original images. Therefore, the HODMD algorithm can also act as an echocardiography video enhancement technique. This fact implies an important contribution to the relatively little research on methods to enhance dynamic ultrasound video sequences, which, in fact, are usually based on supervised learning, rather than on data physics (\citet{liu2025algorithms}). In particular, this work uses the multidimensional iterative HODMD algorithm (\citet{LECLAINCHE2017336}, \citet{Groun2022higher}), which is based on the Higher Order Singular Value Decomposition (HOSVD, \citet{tucker1966some}). As an extension of the SVD algorithm, HOSVD applies this along each spatial resolution to better filter the noise.

The steps of the HODMD algorithm are depicted in Fig.~\ref{fig:hodmd}. The input is a video sequence of echocardiography images, assumed to be homogenized as described in Subsection \ref{sec:cardiac_data_creation}, with dimensions $N_{x} \times N_{y} \times K$, where $N_{x}$ and $N_{y}$ denote the spatial resolution of the frames (or snapshots), and $K$ the number of snapshots. First, each snapshot is reshaped into a vector of dimensions $N_{p} = N_{x} \times N_{y}$, where $N_{p}$ denotes the number of pixels of the snapshot. Then, a tensor is formed from all reshaped snapshots, with dimensions $N_{p} \times K$. Next, the HOSVD algorithm is applied for dimensionality reduction, i.e., to compress this tensor while preserving the predominant spatio-temporal content. Then, a decomposition is performed, obtaining eigenvalues (temporal frequencies) and eigenvectors (representing spatial modes). As an extension of the classical DMD (\citet{schmid2010dynamic}), a number of $d$ consecutive snapshots is considered, improving spectral resolution and enhancing robustness to noise. This is accomplished by applying the higher order Koopman assumption, which relates an echocardiography image with their $d$ previous snapshots with the following expression:

\begin{equation}
\centering
\boldsymbol{V}_{d+1}^{K} \approx \boldsymbol{R}_{1}\boldsymbol{V}_{1}^{K-d} + \boldsymbol{R}_{2}\boldsymbol{V}_{2}^{K-(d-1)} + ... + \boldsymbol{R}_{d}\boldsymbol{V}_{d}^{K-1},
\end{equation}

where $\boldsymbol{V}_{d+1}^{K}$ represents the measurement of the images one time step into the future. It is related with their $d$ previous $K$ equispaced snapshots by the $\boldsymbol{R}_{i}$ Koopman operators. As can be seen, these Koopman operators (\citet{le2017higher}) relate measurements from a non-linear system (in this case, the snapshots) in consecutive time steps with a linear operator of infinite dimension. Therefore, the HODMD algorithm, although formulated with linear operators, as an approximation of the Koopman operator, can deal with the non-linearity which could be present in the echocardiography imaging. Next, the DMD modes $\boldsymbol{u}_{m}$ are computed. The final step consists in a mode expansion process, in which the frequencies, the growth rates, and the amplitudes are obtained with the following expression:

\begin{equation}
\label{eq:mode_exp}
\centering
\boldsymbol{v}(t)\simeq \sum_{m=1}^{M} a_{m}\boldsymbol{u}_{m}e^{\delta_{m}+i\omega_{m}(t-t_{1})} \; \; \textrm{for}\; t_{1} \leq t \leq t_{1}+T, 
\end{equation}

where $\boldsymbol{v}(t)$ represents the spatio-temporal data (here, the input video sequence of echocardiography images) as an expansion of $M$ DMD modes; $t$ is the time, $T$ the sampled timespan, $a_{m}$ the (real) amplitudes, $\boldsymbol{u}_{m}$ the normalized spatial modes, $\delta_{m}$ the growth rates, and $\omega_{m}$ the frequencies.

All these steps previously explained are repeated iteratively until the number of HOSVD modes converges, i.e., remains the same. At the end, a set of DMD modes are obtained, which enable to reconstruct each snapshot from the input sequence. Only the dominant modes are taken for feature extraction and data augmentation, which contain relevant representations of cardiac dynamics, i.e., patterns capturing the behavior associated with the cardiac function. Therefore, these most representative modes are based on the physics of the data. To select these modes, their associated frequencies are compared with those characteristic of the different heart states, previously identified with the HODMD algorithm in \citet{Groun2022higher}. These frequencies characteristic of the heart states are associated with higher amplitudes, and better characterize the temporal evolution of the heart state, the disease progression and, thus, the proximity of the heart failure event. In this way, the modes to be taken are those with higher amplitudes and with frequencies more similar to the standard ones from \citet{Groun2022higher}. The selected modes are also associated with a tolerance, previously configured for the HODMD algorithm. The other modes, which tend to have higher frequencies and smaller amplitudes, are discarded, as not containing patterns of the cardiac dynamics, but the noise of the echocardiography imaging. Both outputs are used to enrich the training database: the reconstructions, and the most representative DMD modes. These allow the deep neural network to focus on clinically meaningful patterns, improving the heart failure prediction performance.

\subsection{Heart Failure Prediction}
\label{sec:heart_fail_pred}

The Heart Failure Prediction stage processes echocardiography images in real-time to predict the times of happening heart failures. The input is specifically a sequence of echocardiography images, adapted to the target acquisition device used in the previous Cardiac Database Creation stage. The output is the estimated age of the corresponding patient in which a heart failure will happen. Therefore, this stage deals with the more challenging and specific task of estimating the concrete time of heart failure, not previously addressed in the related literature to the best of the authors' knowledge. In addition, Masked Autoencoders (MAE) (\citet{he2022masked}) have been adopted in this work, which are more effective and efficient than the modified ViT proposed in \citet{bell2025automatic}. Although being the state of the art in Computer Vision, self-supervised learning (SSL) has rarely been adopted in the related literature on heart failure prediction in echocardiography imaging, even less MAEs. In particular, a joint SSL and supervised learning training scheme is proposed here, allowing the network to learn robust representations even with scarce data, further addressing the usual problem of having a scarce number of samples and reducing computational resources and requirements.          

\begin{figure*}[!ht]

\centering
\includegraphics[width=14 cm]{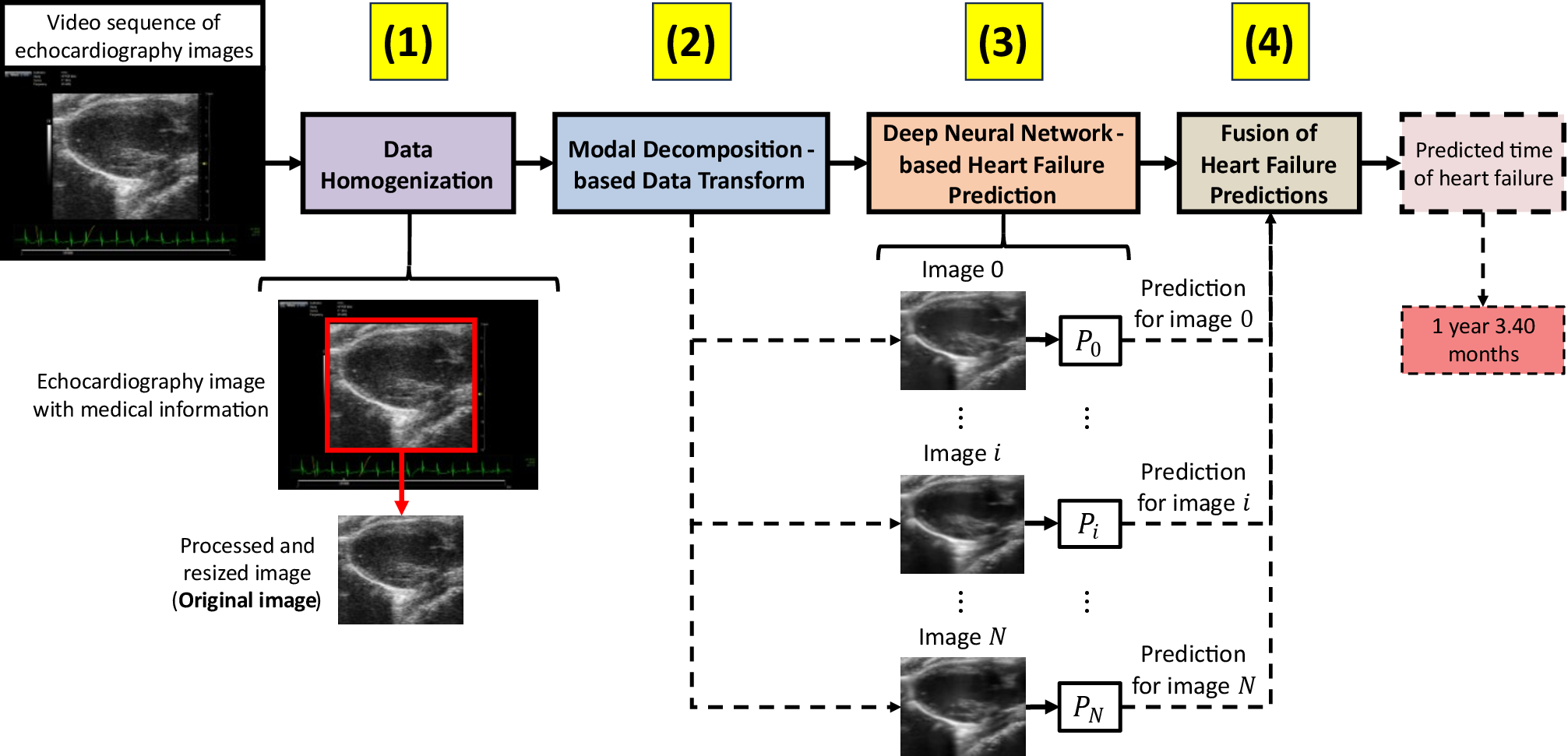}
\caption{Block diagram of the Heart Failure Prediction stage, composed of four phases: (1) Data Homogenization, (2) Modal Decomposition-based Data Transform, (3) Deep Neural Network-based Heart Failure Prediction, and (4) Fusion of Heart Failure Predictions.}
\label{fig:pred}
\end{figure*}

Fig. \ref{fig:pred} shows the Heart Failure Prediction stage, which comprises four phases: (1) Data Homogenization, (2) Modal Decomposition-based Data Transform, (3) Deep Neural Network-based Heart Failure Prediction, and (4) Fusion of Heart Failure Predictions. The first phase, Data Homogenization, adapts the input sequence to the same target acquisition device adopted in the previous Cardiac Database Creation, described in Subsection \ref{sec:cardiac_data_creation}.

The second phase, Modal Decomposition-based Data Transform, obtains images with features which better represent cardiac function and evolution than the original echocardiography imaging, improving the heart failure prediction performance. This is accomplished with the sequential use of the SVD and the HODMD algorithms. Note that this process is analogous to that of the Modal Decomposition-based Data Generation phase described in Subsection \ref{sec:cardiac_data_creation}. However, the images obtained in this phase are used to predict the time of a heart failure from the corresponding sequence, and not for training. Instead, the Modal Decomposition-based Data Generation phase enlarges the training database, using the reconstructions and modes obtained with the SVD and the HODMD algorithms, to address the usual problem in the medicine field of having a limited number of high-quality samples. Experiments testing different types of data obtained with the Modal Decomposition-based Data Transform phase are described in Section \ref{sec:results}: the original echocardiography images (that is, without the Modal Decomposition-based Data Transform), the reconstructions, or modes obtained after applying the SVD or the HODMD algorithm. According to the results (see Section \ref{sec:results}), adopting HODMD data from the test video sequences of echocardiography images as the output of the Modal Decomposition-based Data Transform phase gives better performances. Therefore, the Modal Decomposition techniques adopted in this work effectively capture cardiac dynamics and the evolution of the heart state, and filter out the high noise inherent in the echocardiography imaging: the global effect enables more accurate heart failure time predictions. Moreover, these results also demonstrate the potential of the SVD and HODMD algorithms to broaden the application range from heart disease classification (\citet{bell2025automatic}) to the more challenging and specific task of heart failure time prediction.

In the third phase, Deep Neural Network-based Heart Failure Prediction, each image obtained in the previous phase, Modal Decomposition-based Data Transform, is introduced into a deep neural network to predict a heart failure time. In particular, a ViT is used, trained from scratch with MAEs and a combined scheme of self-supervised (SSL) and supervised learning which improves the representation learning of cardiac dynamics, and so the heart failure time prediction performance, even from scratch and with scarce datasets (\citet{das2024limited}). Therefore, this joint learning scheme further reduces the dependence on large varied databases for a proper training, and so the expensive and time consuming specialist's work involved in their elaboration. To the best of the authors' knowledge, SSL has barely been explored in the related literature on heart failure prediction in echocardiography images, even less Masked Autoencoders (MAE) (\citet{he2022masked}), although being the state of the art in Computer Vision. Moreover, standard architectures have been proposed as the general trend with no specific adaptations nor approaches to address the usual scenario in the medicine field of having a scarce number of high-quality samples. The joint SSL and supervised learning scheme proposed in this work involves two tasks: the Self-supervised Auxiliary Task (SSAT), and the Regression Task. Specifically, the first task, SSAT, aids in the training of the ViT for the heart failure time prediction task (i.e., the Regression Task). This approach combining MAEs and the joint learning scheme is more effective and efficient than the modified ViT from our previous study (\citet{bell2025automatic}), which addresses overfitting to some extent. In addition, this work also incorporates masking, which allows to save considerable computational resources. Moreover, as will be seen in Subsection \ref{sec:analysis_algorithms}, the proposed ViT trained with the MAE scheme and the joint SSL and supervised learning approach achieves a better performance than several existing ViT architectures and Convolutional Neural Networks (CNN), even from scratch.

\begin{figure*}[!ht]

\centering
\includegraphics[width=13.5 cm]{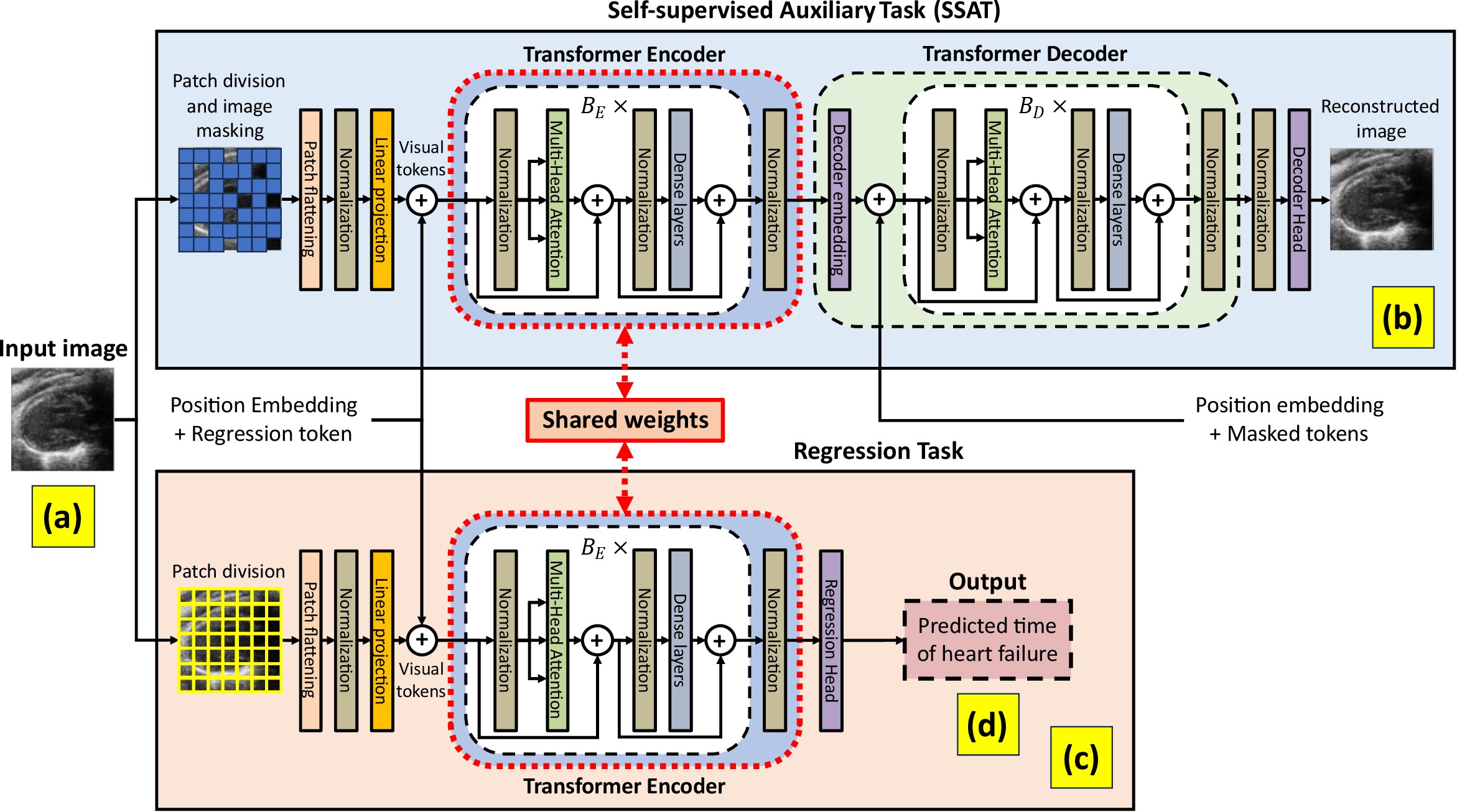}
\caption{Architecture of the proposed deep neural network with the joint self-supervised (SSL) and supervised learning scheme. (a) The input: an image from an echocardiography video sequence, in the form of an original echocardiography sample, a mode, or a reconstruction obtained with the SVD and the HODMD algorithms, result from the Modal Decomposition-based Data Transform phase. (b) The Self-supervised Auxiliary Task (SSAT), aimed to reconstruct the missing patches from the masked image, aiding in training the ViT for the Regression Task. (c) The Regression Task, i.e., the heart failure time prediction task. (d) The predicted heart failure time of the input image.}
\label{fig:model}
\end{figure*}

Fig. \ref{fig:model} shows the architecture of the proposed deep neural network. Two tasks are jointly learned at the same time: the Self-supervised Auxiliary Task (SSAT, Fig. \ref{fig:model} (b)), and the Regression Task (Fig. \ref{fig:model} (c)). The first task, SSAT, is aimed to reconstruct the missing patches from masked images. Its purpose is to improve the representation learning of the deep neural network and to better capture the cardiac dynamics and evolution of the heart state. Thus, for a more accurate heart failure time prediction, i.e., to support the Regression Task (the second). Precisely, the weights of the Transformer Encoders used for both tasks are shared (highlighted in red in Fig. \ref{fig:model}), which then allows for this joint training. Note that any self-supervised learning (SSL) approach can be used, but the Masked Autoencoder (MAE) (\citet{he2022masked}) has been employed due to its popularity and superior performance, becoming the state of the art in Computer Vision (\citet{das2024limited}). This training approach based on joint learning is different to the usual strategy adopted in MAEs (\citet{he2022masked}). That is, the typical approach first performs the pretraining of the ViT for reconstruction of the masked patches, and then the fine-tuning, using the same dataset, for the downstream task, i.e., the final task to address. However, the different training approach here adopted not only improves feature learning of the deep neural network even with limited data, but also allows to speed up the training process by learning both tasks at the same time. Therefore, this joint learning scheme incorporated into the MAE improves training in a more efficient and effective way than the modified ViT from the study in \citet{bell2025automatic}, which addresses overfitting to some extent. The common input for the two tasks is an image representing a heart in a determined state (Fig. \ref{fig:model} (a)). As a result of the Modal Decomposition-based Data Transform phase, this input image can be a reconstruction, or a mode obtained with the SVD and the HODMD algorithms. This fact gives the deep neural network the flexibility of being able to directly use images as input, in the form of original echocardiography data, reconstructions, or modes obtained with the Modal Decomposition techniques. However, as the results in Section \ref{sec:results} will confirm, using data obtained with the HODMD algorithm as input for the deep neural network (and so as the output of the Modal Decomposition-based Data Transform) results in the best performance. The output is a prediction which represents the estimated time in which a heart failure will happen (Fig. \ref{fig:model} (d)). Note that the MAE is in charge of addressing all possible non-linear relationships between the input and the output (in this case, between echocardiography imaging and heart failure times, respectively). Moreover, as a deep neural network, the MAE itself can model arbitrarily complex relationships between the input and the output. However, the combination of the SVD and HODMD algorithms for both feature extraction and data augmentation, and the joint SSL and supervised learning scheme to increase representation learning, even with scarce databases, make the proposed deep neural network able to effectively model these relationships. Note also that, as already explained in \S~\ref{sec:hodmd}, the HODMD algorithm is an approximation of the Koopman operator, so it can deal with the non-linearities existing in the echocardiography imaging.

The processing of the input (Fig. \ref{fig:model} (a)) for the Regression Task (Fig. \ref{fig:model} (c)), for heart failure time prediction (Fig. \ref{fig:model} (d)) closely follows the standard ViT architecture. That is, a division in non-overlapping patches is firstly performed. Next, the spatial dimension of the patches is flattened, followed by a normalization layer and a linear projection. Before introducing the processed patches into the Transformer Encoder, positional information and a regression token, with a role similar to the class token, are added.

The resulting visual tokens are processed by a Transformer Encoder, whose weights are shared with the Transformer Encoder used for the Self-supervised Auxiliary Task (SSAT), and learned at the same time (highlighted in red in Fig. \ref{fig:model}). The configuration followed corresponds to the one of the ViT-T (Tiny) architecture, the shallowest, to minimize overfitting (\citet{das2024limited}). This strategy operates as follows: twelve stacked Transformer blocks (so $B_{E}=12$); each Transformer block includes a multi-head attention layer, with three heads, projection dimension $192$, and without dropout.

An empty branch (skipped connection) and dense layers with two units follow the multi-head attention layer, each one composed of a fully connected layer with the Gaussian Error Linear Unit (GELU) activation function. The skip connections are in charge of addressing the vanishing gradient problem, inherent to deep neural networks, to considerably improve training convergence. The ratio of the dense layers used is $4$, that is, the number of hidden features is four times the amount of input features. The Transformer block introduces another skip connection at the end, which connects the output of the previous one to the output of these dense layers. After all the $B_{E}$ Transformer blocks, features are normalized and introduced into a regression head, whose output is the estimate of the time in which a heart failure will happen (Fig. \ref{fig:model} (d)), according to the input image (Fig. \ref{fig:model} (a)).

Regarding the Self-supervised Auxiliary Task (SSAT, Fig. \ref{fig:model} (b)), the processing is based on the MAE (\citet{he2022masked}). That is, first, patches from the input image (Fig. \ref{fig:model} (a)) are randomly masked before tokenization. After that, the same Transformer Encoder as the one for the Regression Task is used. Precisely, the weights between both Transformer Encoders are shared (highlighted in red in Fig. \ref{fig:model}). Therefore, during training, these weights are adapted to both tasks, improving the representation learning of the ViT used for heart failure time prediction, even with scarce databases. After the $B_{E}$ Transformer blocks, a Transformer Decoder is used for the reconstruction of the missing patches, much shallower than the Encoder. This Decoder has been designed by first incorporating a decoder embedding which adapts the projection dimension to $128$ via a fully connected layer. After adding the masked tokens and the position embedding, two stacked Transformer blocks have been used ($B_{D} = 2$). Each of these blocks includes a multi-head attention layer with $16$ heads. Similarly, empty connections and dense layers with two units and a ratio of 4, each one with the GELU activation function, have also been employed. After all the $B_{D}$ Transformer blocks, features are normalized, the regression token is removed, and a decoder head is used for the reconstruction of the missing patches.

Considering the joint learning on both tasks at the same time, the final convex loss function is as follows:

\begin{equation}
\label{eq:training_loss}
\centering
\textit{L} = (\alpha \times L_{reg}) + [(1 - \alpha) \times L_{SSAT}],
\end{equation}

where $L_{reg}$ is the regression loss between the estimated heart failure time and the real one; $L_{SSAT}$ is the loss component between the original and reconstructed image, computed only for the masked patches (\citet{he2022masked}), and $\alpha$ is the loss scaling factor, weighing the importance of each task. Both loss functions are based on the Mean Square Error (MSE). Note that, after training, like in the usual approach of MAEs (\citet{he2022masked}), for the Deep Neural Network-based Heart Failure Prediction phase, only the branch concerning the Regression Task (Fig.~\ref{fig:model} (c)) is used, and the branch about the Self-supervised Auxiliary Task (SSAT, Fig.~\ref{fig:model} (b)) is discarded.

For optimization, the AdamW algorithm has been used. The values for the momentum and weight decay are 0.9 and 0.05, respectively. On the other hand, the warm-up cosine policy (\citet{Lee2021vision}) has been used to compute the learning rate at iteration $i$ by the expression $\lambda_{i} = 0.5 \times \lambda_{t} \times (1 + cos(\pi \times (i - N_{w}) / (N_{iter} - N_{w})))$, where $N_{iter}$ is the maximum number of iterations. Following the configurations used in \citet{das2024limited}, the target learning rate is fixed to $\lambda_{t} = 2.5 \mathrm{e}{-4}$, and the warmup steps $N_{w}$ to $5$. 

For training, the deep neural network here proposed learns to predict heart failure times using the annotated cardiac database, $\boldsymbol{D}_{train}$, generated according to the Cardiac Database Creation stage described in Subsection \ref{sec:cardiac_data_creation}. The adopted database splitting scheme takes $60 \% - 20 \%$ for training and validation, respectively, and balances the number of samples from each heart state in each set to have similar distributions of heart failure times. Regarding the batch generation scheme, the size has been set to 64 images, because of the limitations of the physical memory of the GPUs available (Nvidia Tesla A100 and Nvidia RTX A4500). This includes data augmentation techniques based on geometric transformations (in particular, resizing, random horizontal flips, and random erasing) to increase robustness to the variation of the perspective of the heart, and to marginally improve final performance beyond the Modal Decomposition techniques.    

\begin{figure*}[!ht]

\centering

\includegraphics[width=8 cm]{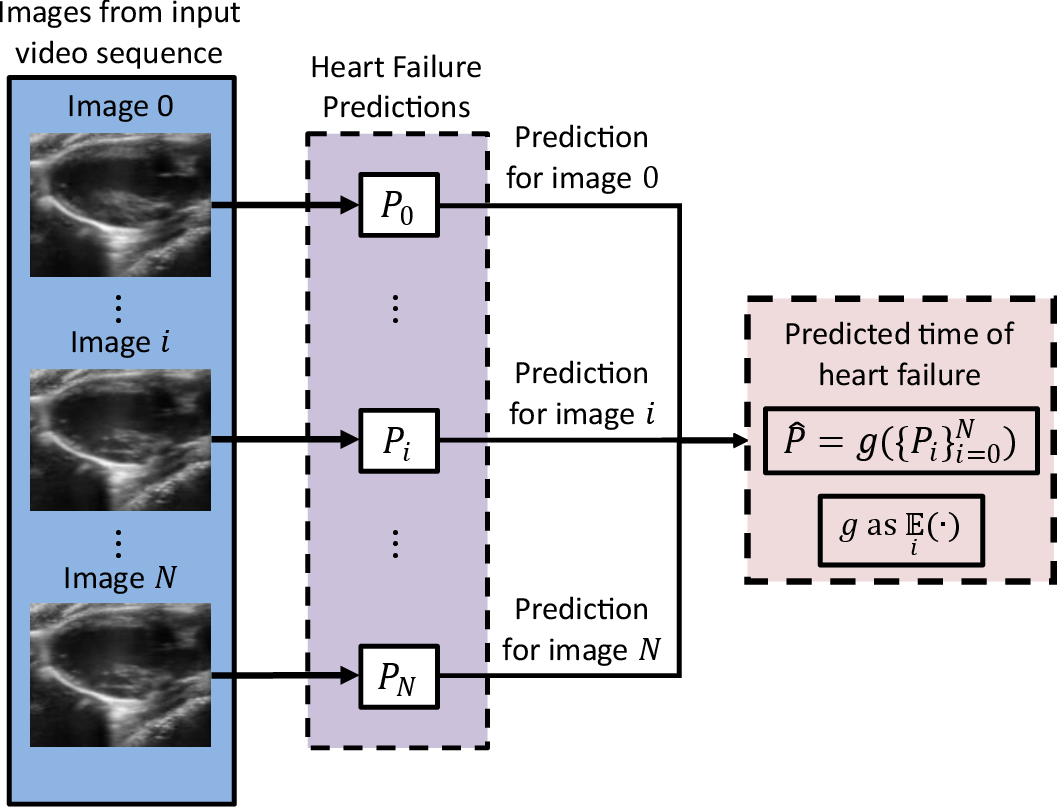}
\caption{Illustration of the Fusion of Heart Failure Predictions phase. The heart failure time predictions associated to each image $i$, originating from a test sequence, are fused, and the average value determines the estimated time in which a heart failure will happen.}
\label{fig:fusion}
\end{figure*}

The fourth and last phase of the framework, Fusion of Heart Failure Predictions (cf. Fig. \ref{fig:fusion}), an estimate of the time of happening a heart failure in the input test sequence is obtained from the combination of multiple predictions. Each image $i$ from the $N+1$ images depicted was obtained from the input test sequence according to the process explained in the second phase, Modal Decomposition-based Data Transform. That is, these images can be either original echocardiography images, or modes or reconstructions obtained with the SVD and HODMD modes. Then, each of these images is processed by the deep neural network as explained in the third phase, Deep Neural Network-based Heart Failure Prediction. Therefore, the deep neural network yields for each of these images has a time prediction $P_{i}$, that is, the estimate of the age of the patient in which a heart failure will happen. The set of time predictions $\{P_{i}\}$ is then used to compute a unique failure time prediction associated with the input test sequence $\hat{P}$. In particular, the average of that set is calculated. As a result, an estimate of the time in which a heart failure will happen is obtained. Note that relying on a sequence of echocardiography images instead of a single image enables a more robust heart failure time prediction, as more data can help health professionals to make more confident and accurate heart prognoses. Moreover, certain heart pathologies might show characteristic patterns in the heartbeats which, therefore, constitute dynamic information which can be obtained from sequences, and which could be valuable to accurately predict heart failures. 

\section{Database}
\label{sec:database}

\begin{figure*}[!ht]

\centering
\includegraphics[width=10 cm]{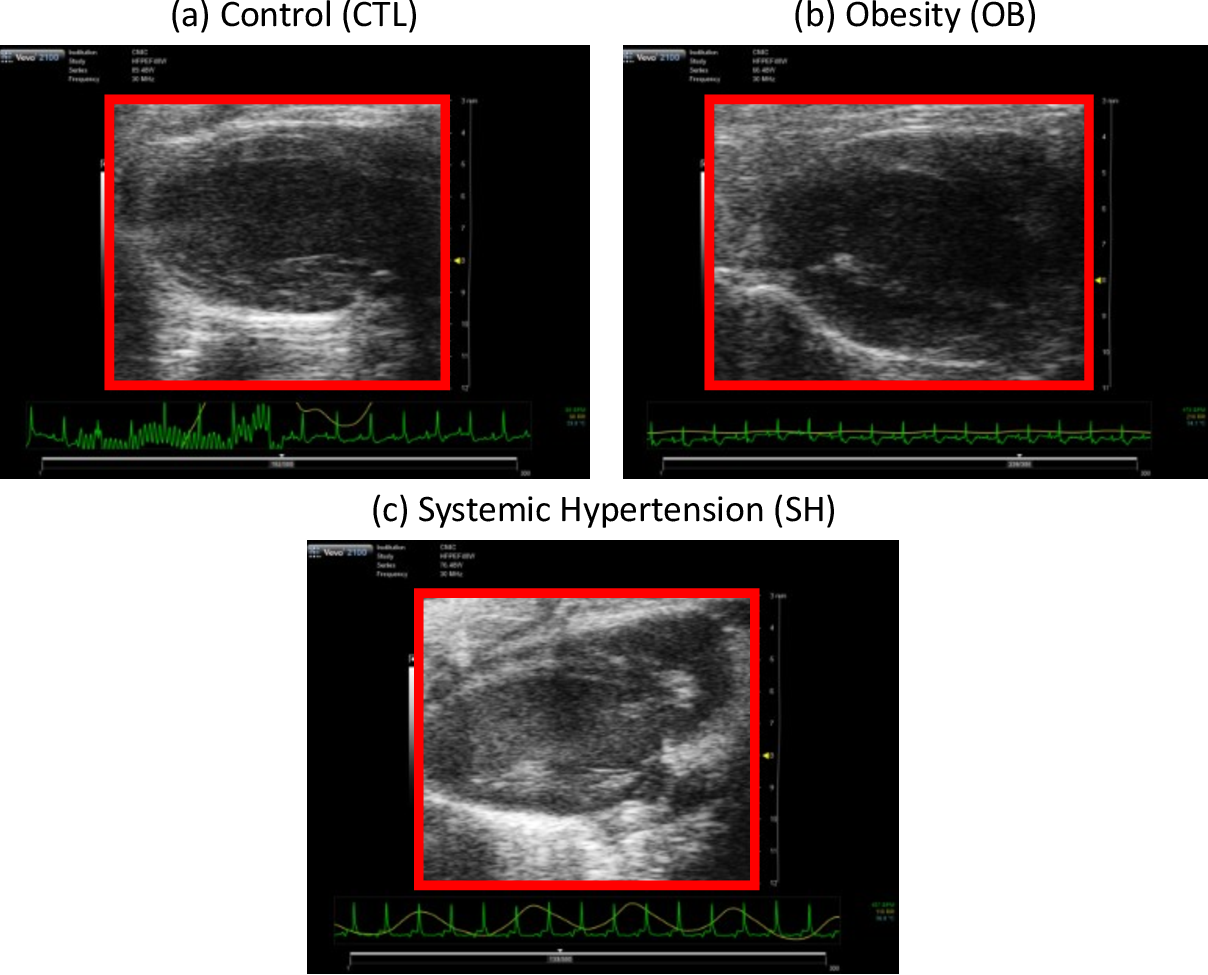}
\caption{Sample images of the database of echocardiography imaging, showing the studied heart states. The regions of interest (ROI) are delimited by the red boundaries.}
\label{fig:database}
\end{figure*}

A database of video sequences of echocardiography images has been used to test and validate the proposed heart failure prediction system. This database has been elaborated in collaboration with the Centro Nacional de Investigaciones Cardiovasculares (CNIC). Each video sequence represents a heart in a determined state, and has been provided together with an annotation of the time of a heart failure. In particular, the following heart states have been studied: Control (CTL), Obesity (OB), and Systemic Hypertension (SH). A sample image from each of these heart states is shown in Fig.~\ref{fig:database}. As anticipated, echocardiography imaging is inherently afflicted by noise. In addition, image acquisition has been performed from different perspectives, further increasing the complexity of the heart failure prediction task. As specified in the Data Homogenization phase described in Subsection \ref{sec:cardiac_data_creation}, only the area of the heart, i.e., the region of interest (ROI), is considered for the rest of the system. The obtained heart areas, however, have different resolutions, demonstrating the heterogeneity of the echocardiography imaging. Specifically, in mean and standard deviation, the spatial dimensions are $(675.68 \pm 47.28) \times (583.15 \pm 0.49)$ pixels. Nevertheless, the aspect factor among the images is $1.16 \pm 0.08$, meaning that it barely changes and is close to 1, so the heart areas are practically square. Therefore, resizing to square images barely deforms the heart areas and does not suppose a degradation of the heart failure prediction performance, as preserving the features associated with cardiac function and disease progression over time.  

\begin{table}[!ht]
\centering
\caption{Overview of the main characteristics of the cardiac database.}
\label{tab:lax_data}
\resizebox{\textwidth}{!}{%
\begin{tabular}{@{}ccccccc@{}}
\toprule
Heart State &
  \begin{tabular}[c]{@{}c@{}}Age of heart failure\\ (months) \end{tabular} &
  Set &
  \# Sequences &
  \# Snapshots &
  \# SVD Modes &
  \# HODMD Modes \\ \midrule
\multirow{3}{*}{\begin{tabular}[c]{@{}c@{}}Control\\ (CTL)\end{tabular}} &
  \multirow{3}{*}{\begin{tabular}[c]{@{}c@{}} $23.86 \pm 5.23$ \end{tabular}} &
  Training &
  35 &
  \num[group-separator={~}]{10204} &
  175 &
  1334 \\
    &     & Validation    & 12  & 3496                             & 60  & 457  \\
    &     & Test          & 9   & 2619                             & 45  & 321  \\
\multirow{3}{*}{\begin{tabular}[c]{@{}c@{}}Obesity\\ (OB)\end{tabular}} &
  \multirow{3}{*}{\begin{tabular}[c]{@{}c@{}} $22.03 \pm 6.35$ \end{tabular}} &
  Training &
  28 &
  8170 &
  140 &
  1140 \\
    &     & Validation    & 9   & 2700                             & 45  & 369  \\
    &     & Test          & 11  & 3194                             & 55  & 412  \\
\multirow{3}{*}{\begin{tabular}[c]{@{}c@{}}Systemic Hypertension\\ (SH)\end{tabular}} &
  \multirow{3}{*}{\begin{tabular}[c]{@{}c@{}} $20.77 \pm 6.65$ \end{tabular}} &
  Training &
  31 &
  8919 &
  155 &
  1186 \\
    &     & Validation    & 12  & 3136                             & 60  & 422  \\
    &     & Test          & 10  & 3000                             & 50  & 380  \\ \midrule
\multicolumn{3}{c}{\textbf{Total}} & 157 & \num[group-separator={~}]{45438} & 785 & 6021 \\ \bottomrule
\end{tabular}%
}
\end{table}

Table~\ref{tab:lax_data} summarizes the main characteristics of the cardiac database used in this work. Considering the annotation of the time of heart failure provided for each video sequence, an average and a standard deviation of the age of this heart failure event has been represented for each state. As expected, unhealthy classes (OB, and SH) tend to have earlier heart failures than in the case of healthy hearts (CTL). This cardiac database is generated according to the process carried out in the Modal Decomposition-based Data Generation phase, described in Subsection \ref{sec:cardiac_data_creation} (see Fig. \ref{fig:cardiac_data_creation}). That is, the SVD-based data were obtained by applying the SVD algorithm on each original video sequence of echocardiography images. In this way, for each of these sequences, reconstructions and modes were obtained with the SVD algorithm (denoted as $\boldsymbol{D}_{SVD1\_R}$ and $\boldsymbol{D}_{SVD1\_M}$, respectively, following the notation from Subsection \ref{sec:cardiac_data_creation}). After that, the HODMD-based data were generated by applying the HODMD algorithm on each sequence, but using their associated SVD reconstructions instead of the original images. As a result, for each sequence, reconstructions and modes were obtained ($\boldsymbol{D}_{HODMD\_R}$, $\boldsymbol{D}_{HODMD\_M}$, respectively). Finally, the SVD algorithm was applied to the HODMD modes of each sequence. In this case, only the reconstructions of the HODMD modes were taken ($\boldsymbol{D}_{SVD2\_R}$). In Table~\ref{tab:lax_data}, the column '\# Snapshots' indicates the total number of original snapshots of the sequences. Note that this number equals the number of associated reconstructed ones after applying the SVD algorithm, and also the number of reconstructions obtained after using the HODMD algorithm. The columns '\# SVD modes' and '\# HODMD modes' represent the total number of SVD modes and of HODMD modes, respectively.

\begin{table}[!ht]
\centering
\caption{Summary of the combinations of types of data to form $\boldsymbol{D}_{train}$, used in the Cardiac Database Creation stage.}
\label{tab:training_cases}
\resizebox{\textwidth}{!}{
\begin{tabular}{@{}cccccccc@{}}
\cmidrule(l){3-8}
   &          & \multicolumn{2}{c}{SVD-based Data} & \multicolumn{2}{c}{HODMD-based Data} & SVD (DMD Modes) & \multirow{2}{*}{\# Training Samples} \\ \cmidrule(r){1-7}
Case & $\boldsymbol{D}_{orig}$ & $\boldsymbol{D}_{SVD1\_R}$ & $\boldsymbol{D}_{SVD1\_M}$    & $\boldsymbol{D}_{HODMD\_R}$ & $\boldsymbol{D}_{HODMD\_M}$    & $\boldsymbol{D}_{SVD2\_R}$ &                                  \\ \midrule
1    & $\times$ &               &          &               &          &               & \num[group-separator={~}]{27293} \\
2    &          & $\times$      &          &               &          &               & \num[group-separator={~}]{27293} \\
3    & $\times$ & $\times$      & $\times$ &               &          &               & \num[group-separator={~}]{55056} \\
4    & $\times$ & $\times$      &          &               &          &               & \num[group-separator={~}]{54586} \\
5    &          & $\times$      & $\times$ &               &          &               & \num[group-separator={~}]{27763} \\
6    &          &               &          & $\times$      &          &               & \num[group-separator={~}]{27293} \\
7    &          &               &          & $\times$      & $\times$ &               & \num[group-separator={~}]{38273} \\
8    &          & $\times$      &          & $\times$      &          &               & \num[group-separator={~}]{54586} \\
9    & $\times$ & $\times$      &          & $\times$      &          &               & \num[group-separator={~}]{81879} \\
10   &          &               &          & $\times$      & $\times$ & $\times$      & \num[group-separator={~}]{49253} \\
11   &          & $\times$      & $\times$ & $\times$      & $\times$ &               & \num[group-separator={~}]{66036} \\
12   &          & $\times$      & $\times$ & $\times$      & $\times$ & $\times$      & \num[group-separator={~}]{77016} \\
13   & $\times$ & $\times$      & $\times$ & $\times$      & $\times$ &               & \num[group-separator={~}]{93329} \\
14 & $\times$ & $\times$         & $\times$        & $\times$          & $\times$         & $\times$        & \num[group-separator={~}]{104309}    \\ \bottomrule
\end{tabular}%
}
\end{table}

Table~\ref{tab:training_cases} shows the combinations of the types of data used to form the training database $\boldsymbol{D}_{train}$, organized as training cases. Experiments using different types of test data, resulting from the Modal Decomposition-based Data Transform phase described in Subsection \ref{sec:heart_fail_pred}, have also been performed. In this way, the impact of each type of data on the final performance, as well as the robustness to noise, could be evaluated. Note that the absolute value, the real, and the imaginary parts of the HODMD modes have been taken to increase the number of samples (i.e., images) of the database and so to obtain the total number of training samples in each case. Also note that the last column, i.e., SVD (HODMD Modes), refers to the reconstructions of the HODMD modes obtained with the SVD algorithm (that is, $\boldsymbol{D}_{SVD2\_R}$, according to the notation presented in Subsection \ref{sec:cardiac_data_creation}). Similarly, the absolute value, and the real and imaginary parts have been taken to increase the database.

\section{Results}
\label{sec:results}

The proposed system has been tested with cardiac datasets built from $\boldsymbol{D}_{orig}$, using the procedure described in Fig. \ref{fig:cardiac_data_creation} and in Subsection \ref{sec:cardiac_data_creation}, according to the combinations in Table \ref{tab:training_cases}. Additionally, the strategy here proposed has been compared with established ViT and CNN-based algorithms, including DeiT (\citet{touvron2021training}), ResNet50-version 2 (\citet{he2016identity}), and Swin-version 2 (\citet{liu2022swin}).

The following metrics have been utilized to measure the performance of each algorithm in terms of heart failure prediction accuracy and computational cost: mean ($\mu$) and standard deviation ($\sigma$) of the predicted age of heart failure, estimation error, the root mean squared error (\textit{RMSE}, or error margin), maximum and minimum errors, estimated floating point operations per second (GFLOPs, or Gigaflops), and average processing time per image $\bar{t}$.

The mean ($\mu$) and standard deviation ($\sigma$) of the predicted age of heart failure are calculated for each heart state using the computed values of heart failure times of all the corresponding test video sequences. That is, given a set of predicted times of heart failures corresponding to the test video sequences of a determined heart state $y$, $\{\Hat{P}_{j}\}$, the mean and the standard deviations for this heart state are calculated based on the set as follows:

\begin{equation}
\centering
\begin{split}
\mu = \mathbb{E}\big[\{\hat{P}_{j}\}\big]; \\
\\
\sigma = \sqrt{\mathbb{E}\big[(\{\hat{P}_{j}\} - \mu)^{2}}\big].
\end{split}
\end{equation}

The estimation error in a test sequence $j$ is the difference between the associated predicted time of heart failure $\Hat{P}_{j}$ and the true time ${T}_{j}$, expressed as $\Hat{P}_{j} - {T}_{j}$. Derived from this, the root mean squared error (\textit{RMSE}, or error margin from now on), and the maximum and minimum errors are computed for each heart state as follows:

\begin{equation}
\centering
\begin{split}
\textit{RMSE} = \sqrt{\mathbb{E}\big[(\{\hat{P}_{j}\} - \{T_{j}\})^{2}}\big]. \\
\textit{Max error} = \max{\big(\big\{\{\hat{P}_{j}\} - \{T_{j}\}\big\}\big)}. \\
\textit{Min error} = \min{\big(\big\{\{\hat{P}_{j}\} - \{T_{j}\}\big\}\big)}.
\end{split}
\end{equation}

The maximum and minimum errors have been represented with and without sign (that is, in absolute value). For brevity, error with sign is denoted as (w/), and error without sign as (w/o). Therefore, a negative error, i.e., $\Hat{P}_{j} - {T}_{j} < 0$, would indicate that a predicted heart failure would happen sooner than the real time of heart failure, and it would be later otherwise. Unless explicitly indicated otherwise, heart failure times are represented in months.

The average processing time per sample $\bar{t}$ represents the average time taken by the proposed system to process a sample (either image or sequence) and get the heart failure prediction. It comprises the SVD algorithm ($\bar{t}_{SVD}$), the HODMD algorithm, including the HOSVD dimensionality reduction ($\bar{t}_{HOSVD}$) prior to the HODMD itself ($\bar{t}_{HODMD}$) (\citet{Bell2023HODMD}), and the Deep Neural Network-based Heart Failure Prediction phase ($\bar{t}_{pred}$). The throughput is measured in frames per second, fps, and represents the number of snapshots which can be processed by the deep neural network in a second. Therefore, this metric assesses the capability of the deep neural network for real-time applications. The throughput is calculated as $1/\bar{t}_{pred}$. Unless explicitly specified otherwise, the times which refer to the computational cost are measured in milliseconds.

The experiments have been performed using a cluster with Intel Xeon Gold 6240R, AMD Ryzen Threadripper PRO 5995WX, three Tesla A100 GPU, and four Nvidia RTX A4500 working in parallel for training, and Intel Xeon Gold 6230 and one Tesla V100 GPU for testing. These resources allow to manage the high amount of data conforming the created cardiac database, and also to speed up and help training convergence.

The configurations of the SVD and HODMD algorithms, the deep neural network, and of the prediction are introduced next.

Regarding the Modal Decomposition-based Data Generation and Transform phases, the configuration is as follows. In the SVD algorithm, the number of retained modes is $5$, which was found to provide fair reconstructions and also to effectively filter noise, according to the contribution of the modes in the frames. Regarding the HODMD algorithm, the minimum number of snapshots a sequence must have to be able to apply the HODMD algorithm has been set to $100$, following \citet{Groun2022higher}. This ensures to cover at least three heart cycles, considered enough to obtain fair reconstructions and to adequately capture cardiac dynamics and the evolution of the heart state. Moreover, the number of snapshots taken from each sequence $K$ has been set to the total number of frames of the corresponding sequence. In this way, the maximum number of cardiac cycles possible can be captured and the cardiac and respiratory frequencies can be more accurately estimated, leading to the extraction of representative modes characterizing the physics of the data. The time interval between snapshots $\Delta t = 4~ms$ and the tolerances $\epsilon_{SVD} = \epsilon_{DMD} = 5 \mathrm{e}{-4}$ have been selected according to \citet{Groun2022higher}: on the one hand, the value of $\Delta t$ is determined according to the ultrasound scanner configuration performed by a specialist, to ensure proper extraction of the cardiac and respiratory frequencies. On the other hand, the chosen tolerances for the dimensionality reduction and amplitude truncation steps, respectively, are greater than the noise level, and lead to a reasonable number of characteristic frequencies and modes describing the cardiac dynamics and evolution of the heart state over time. The index $d$ was set individually for each sequence to achieve fair reconstructions of the snapshots: in particular, $d = \left \lfloor K/5 \right \rfloor$ results adequate and consistent with the calibration recommendations of \citet{le2017higher}. Moreover, alternative choices of $d$ lead to reconstructions with severe distortions and artifacts which fail to reproduce the cardiac dynamics and heart function. Therefore, erroneous features are extracted in those cases.    

For the Deep Neural Network-based Heart Failure Prediction phase, input images have been resized to $224 \times 224$, and are divided into non-overlapping patches of $16 \times 16$ pixels. A random masking ratio of $0.75$ has been applied to the patches in the Self-supervised Auxiliary Task (SSAT). This allows a sufficient number of patches for a proper training, given the available GPU memory. Regarding the deep neural network, as already described in Subsection \ref{sec:heart_fail_pred}, the ViT architecture corresponds to the shallowest configuration, which is ViT-T (Tiny) (\citet{das2024limited}). Indeed, deeper configurations of the architecture were intentionally discarded, because these would be too complex and more prone to overfitting, especially in view of the available GPU memory and the actual scenario of scarce number of samples. Regarding the decoder configuration from the MAE, here we follow the standard setup from \citet{he2022masked}. The scaling factor to weight the training loss components was set to $\alpha = 0.1$, in line with \citet{das2024limited}. This emphasizes the representation learning of cardiac dynamics from the echocardiography data (see Eq. \ref{eq:training_loss} in Subsection \ref{sec:heart_fail_pred}). 

Finally, for the Fusion of Heart Failure Predictions phase, the global predicted time of happening a heart failure for a given test sequence is computed with the average of the predicted time values associated to the corresponding snapshots.

\subsection{Performance evaluation using different types of data}
\label{sec:analysis_data}

Here, we discuss the performance in terms of heart failure prediction accuracy for different training cases and types of data used for test (i.e., test data obtained with the Modal Decomposition-based Data Transform phase described in Subsection \ref{sec:heart_fail_pred}). First, the influence of the types of data used in training and testing on the final performance is assessed. Then, the heart failure prediction performance in each heart state obtained with the best configuration is presented.

\begin{table*}[!ht]
\centering
\caption{Error margin results comparing different types of data for test and training cases using the proposed framework (best results in \textbf{bold}, in Case 14). Refer to Table~\ref{tab:training_cases} for the specifications of each training case.}
\label{tab:results_system_error_margin}
\resizebox{\textwidth}{!}{%
\begin{tabular}{@{}cccccccccccccccc@{}}
\toprule
                       &                                        & \multicolumn{14}{c}{Error margin $\downarrow$}                                                                                                                                                                                                                                                                                                                                                                                                                                                                                                                                                                                                                                                                                                                                                       \\ \midrule
                       &                                        & Case 1                                               & Case 2                                               & Case 3                                               & Case 4                                               & Case 5                                               & Case 6                                               & Case 7                                               & Case 8                                               & Case 9                                               & Case 10                                              & Case 11                                              & Case 12                                              & Case 13                                              & Case 14                                               \\ \cmidrule(l){3-16} 
                       & Original                               & 4.75                                                 & 5.05                                                 & 4.67                                                 & 5.12                                                 & 4.58                                                 & 5.53                                                 & 4.57                                                 & 5.55                                                 & 4.78                                                 & 4.65                                                 & 4.59                                                 & 4.63                                                 & 4.56                                                 & $\mathbf{4.54}$                                                  \\
\multirow{2}{*}{SVD}   & Reconstructions                        & 5.09                                                 & 5.30                                                 & 4.70                                                 & 5.19                                                 & 4.59                                                 & 5.80                                                 & 4.57                                                 & 5.62                                                 & 4.90                                                 & 4.65                                                 & 4.60                                                 & 4.64                                                 & 4.56                                                 & 4.55                                                  \\
                       & Modes                                  & 7.08                                                 & 7.23                                                 & 4.72                                                 & 5.58                                                 & 4.75                                                 & 6.73                                                 & 4.71                                                 & 7.29                                                 & 6.71                                                 & 4.70                                                 & 4.69                                                 & 4.72                                                 & 4.66                                                 & 4.61                                                  \\
\multirow{2}{*}{HODMD} & Reconstructions                        & 5.10                                                 & 5.23                                                 & 4.69                                                 & 5.18                                                 & 4.59                                                 & 5.78                                                 & 4.57                                                 & 5.60                                                 & 4.89                                                 & 4.65                                                 & 4.60                                                 & 4.64                                                 & 4.56                                                 & 4.55                                                  \\
                       & Modes (abs)                            & 4.87                                                 & 4.55                                                 & 6.01                                                 & 4.72                                                 & 4.67                                                 & 4.80                                                 & 4.84                                                 & 5.05                                                 & 5.33                                                 & 4.59                                                 & 4.75                                                 & 4.65                                                 & 4.71                                                 & $\mathbf{4.53}$                                       \\ \cmidrule(l){2-16} 
\multicolumn{1}{l}{}   & \multicolumn{1}{l}{\# Samples (train)} & \multicolumn{1}{l}{\num[group-separator={~}]{27293}} & \multicolumn{1}{l}{\num[group-separator={~}]{27293}} & \multicolumn{1}{l}{\num[group-separator={~}]{55056}} & \multicolumn{1}{l}{\num[group-separator={~}]{54586}} & \multicolumn{1}{l}{\num[group-separator={~}]{27763}} & \multicolumn{1}{l}{\num[group-separator={~}]{27293}} & \multicolumn{1}{l}{\num[group-separator={~}]{38273}} & \multicolumn{1}{l}{\num[group-separator={~}]{54586}} & \multicolumn{1}{l}{\num[group-separator={~}]{81879}} & \multicolumn{1}{l}{\num[group-separator={~}]{49253}} & \multicolumn{1}{l}{\num[group-separator={~}]{66036}} & \multicolumn{1}{l}{\num[group-separator={~}]{77016}} & \multicolumn{1}{l}{\num[group-separator={~}]{93329}} & \multicolumn{1}{l}{\num[group-separator={~}]{104309}} \\ \bottomrule
\end{tabular}%
}
\end{table*}

Table \ref{tab:results_system_error_margin} reveals the impact of using the different types of data for testing and of the training cases presented in Table \ref{tab:training_cases} on the heart failure time prediction performance. In this table, the error margin has been represented. The training cases lead to a general improvement, because of the decreasing error margin values, with respect to the case 1 (with only original images), showing the usefulness of the presented training cases, and so the benefit of incorporating more training samples, as expected. Precisely, the case 14 which considers all types of data (original images and both SVD and HODMD modes and reconstructions) leads to the best error margin: $4.54$ months. Note that, in this case 14, the performance barely changes regardless of the type of testing data used, which demonstrates the robustness of the proposed system to noise and to different types of data. This also demonstrates the efficacy of the extension of the database creation procedure from \citet{bell2025automatic} in this work, as a larger annotated database is generated. Specifically, this database incorporates the original echocardiography images, enabling extended cases, instead of a subset of the data generated. Also note that the training cases affect the variation of the performance. To illustrate this fact, let's compare the cases 9 and 11. The case 9 uses original images and reconstructions from both the SVD and the HODMD algorithms, whereas the case 11 incorporates modes instead of the original images. Therefore, the case 9 has more samples than the case 11, but leads to a lower performance. Also note the consistent improvement produced in the training cases which incorporate HODMD modes, and also in those training cases which include the SVD reconstructions of the HODMD modes, which are the cases 7, 10, 11, 12, 13, and 14. Therefore, these two observations reveal the relevance of HODMD modes. This is due to that the HODMD modes describe the dynamics of the data and contain the temporal information useful to make the proposed ViT to effectively learn cardiac dynamics and the evolution of the heart state; this leads in turn to more accurate predictions of the heart failure time. In addition, the data generated with the SVD and the HODMD algorithms have less noise than the original data (inherently with much noise), once more, leading to more representative features. This shows the importance of using the SVD and the HODMD algorithms for both data augmentation and extraction of features more representative of cardiac function and the evolution of the heart state over time, allowing more accurate heart failure prediction times. The SVD and the HODMD algorithms also improve the capability of the proposed ViT to estimate the heart failure time in the usual scenario of having a scarce number of samples. In turn, this also demonstrates the usefulness of the Modal Decomposition techniques to deal with the difficulty to elaborate varied databases with high quality enough in the medical field in both costs and very hard specialized work. In summary, the results show the potential of the data creation procedure proposed in this work, broadening the application range from heart disease classification (\citet{bell2025automatic}) to the more challenging and specific task of heart failure time prediction. The results also show the effectiveness of the SVD and of the HODMD algorithms as echocardiography video enhancement techniques, dealing with the high amount of noise inherent to the original echocardiography images, leading to performance improvements, and making the proposed system more robust to noise. Precisely, the SVD and the HODMD algorithms would also effectively filter other types of noise and would also generate features better representing the heart function. Therefore, these Modal Decomposition techniques suppose an important contribution to the relatively little research on methods to enhance dynamic ultrasound video sequences (\citet{liu2025algorithms}), usually based on supervised learning, unlike the SVD and the HODMD algorithms, based on data physics. 

Table \ref{tab:results_per_class} presents the heart failure prediction results for each heart state with the best configuration inferred from Table \ref{tab:results_system_error_margin}: training case 14 and original images as test data. This comparison is informative because of having more original images than modes, leading to more reliable prediction results in the fusion of the estimated heart failure times. The performance is balanced among the considered heart states, also considering the similar number of test samples in terms of both images and sequences in these states. A general error margin of $4.54$ months is reached, which is accurate enough for medical applications considering the scarce number of original training samples (images) for each class for a proper training of deep neural networks (which usually require hundreds of thousands or even millions of samples).

These results suggest that the ability of the proposed deep neural network to capture cardiac dynamics and to predict heart failures would be enhanced with a larger database. However, collecting large varied high-quality databases of echocardiography imaging requires high costs, effort, and medical expertise. Finally, observe that the means ($\mu$) and standard deviations ($\sigma$) of the real and predicted ages of heart failures are similar. This means that the proposed system effectively reproduces the statistics of the real ages of heart failures under the different cardiac conditions. Therefore, this consistency demonstrates the capability of the proposed deep neural network to model all possible complex non-linear relationships between echocardiography imaging and heart failure times.

\subsection{Comparison with Alternative Algorithms}
\label{sec:analysis_algorithms}

This second part of the experimental analysis shows the evaluation of the proposed system against other alternative deep neural network architectures. This is, the proposed ViT trained with the joint SSL and supervised learning scheme is replaced with other well established deep neural networks, while the rest of the framework remains the same. Specifically, the compared architectures are DeiT (\citet{touvron2021training}), ResNet50-version 2 (\citet{he2016identity}), and Swin-version 2 (\citet{liu2022swin}). These deep neural networks have been selected for the comparisons to cover a range of competitive CNNs and ViTs which introduce alternative mechanisms for better training regarding the architecture and adoption of auxiliary databases. First, the configuration of the proposed system and the adaptations required for each alternative deep neural network are described. Next, the configuration and the necessary adaptations of the alternative algorithms are described. Finally, a comparative performance analysis is presented between the proposed ViT and the other deep neural networks. In this case, the most representative training cases (1, 7, 10, 11, 12, 13, 14) and the main types of test data (Original, HODMD-based reconstructions, and HODMD modes) are considered.

\begin{table*}[!ht]
\centering
\caption{Heart failure time prediction performance for the studied heart states using $\mu$, $\sigma$, error margin, maximum and minimum errors (with: w/, and without sign: w/o) obtained with the best configuration of the proposed framework (training case 14 and original images as test data. Refer to Table~\ref{tab:training_cases} for the specifications of each training case).}
\label{tab:results_per_class}
\resizebox{\textwidth}{!}{%
\begin{tabular}{@{}ccccccccc@{}}
\toprule
\multirow{2}{*}{Heart State} & \multicolumn{2}{c}{Age of heart failure ($\mu \pm \sigma$)} &              &                 & \multicolumn{2}{c}{\textit{Min error} $\left| \downarrow \right|$} & \multicolumn{2}{c}{\# Test samples} \\ \cmidrule(l){2-9} 
                             & Real                & Predicted          & Error margin $\downarrow$ & \textit{Max error} (w/) $\left| \downarrow \right|$ & (w/)           & (w/o)        & Images       & Sequences       \\ \midrule
Control (CTL)                & $27.83 \pm 0.00$    & $23.52 \pm 0.91$   & 4.41         & - 2.81          & - 6.29         & 2.81         & 2619         & 9               \\
Obesity (OB)                 & $23.50 \pm 4.20$    & $23.79 \pm 0.96$   & 4.40         & 6.95            & - 4.59         & 1.48         & 3194         & 11              \\
Systemic Hypertension (SH)   & $19.55 \pm 3.38$    & $22.80 \pm 1.26$   & 4.81         & 9.62            & - 1.55         & 0.52         & 3000         & 10              \\ \midrule
\textbf{Total}                        & $23.48 \pm 4.59$    & $23.38 \pm 1.14$   & 4.54         & 9.62            & - 6.29         & 0.52         & 8813         & 30              \\ \bottomrule
\end{tabular}%
}
\end{table*}

\begin{enumerate}

\item Configuration and adaptation of the alternative algorithms: The same configuration already described in Subsection \ref{sec:analysis_data} applies to the SVD and HODMD algorithms, the proposed ViT, and to the prediction. 


Regarding the alternative deep neural networks, i.e., the DeiT, ResNet50-v2, and Swin-v2, several implementation considerations must be addressed before introducing the configurations and parameters selected. First, these networks originally process three-channel inputs (RGB images), but the echocardiography data used in this work comprise one single channel. In addition, these networks were originally designed for multi-class classification instead of for regression problems, for example, for the ImageNet challenge classes (\citet{russakovsky2015imagenet}). Consequently, adjustments in the architectures were necessary to adapt the alternative deep neural networks to both one-channel data and regression tasks (i.e., to heart failure time prediction). On the one hand, to address the channel differences, each input sample has been replicated across three channels and concatenated along the channel dimension before being fed into the networks. On the other hand, to enable a regression output, the final classification layer of each architecture has been replaced with a linear layer whose output is a single continuous value representing the estimated time of happening a heart failure. The alternative architectures have been trained from scratch, with random initialization of their weights, to assess the efficacy of the proposed joint SSL and supervised learning strategy. However, in the specific case of ResNet50-v2, two initialization strategies have been tested: random weights, and transfer learning using weights pretrained on the ImageNet dataset (\citet{russakovsky2015imagenet}). The latter supposes a favorable starting point, because this initialization potentially speeds up training convergence and enhances generalization after fine-tuning on the cardiac database. This means that ResNet50-v2 also benefits from the auxiliary database, unlike the proposed ViT and the other deep neural networks, which are trained only with the cardiac database, thus implying an additional advantage. Unlike the proposed ViT trained with the joint SSL and supervised learning scheme, the alternative architectures (i.e., DeiT, Swin-v2, and ResNet50-v2) adopt mixup for training, in addition to the conventional data augmentation techniques and the modal decomposition to augment the databases, as improving performance. On the contrary, mixup was discarded for the training of the proposed ViT because, according to preliminary experiments, this model shows a performance degradation. This is likely because mixup interferences with the feature representation learning made by the MAE. 

All the compared deep neural networks, except for Swin-v2, use input images resized to $224 \times 224$. In the case of Swin-v2, the input image size has been set to $256 \times 256$. In this way, the default configurations of the input size for these architectures are followed, as well as the constraints of the GPU memory for training. The patch size in DeiT has been set to $16$, whereas the window size in Swin-v2 has also been fixed to $16$ pixels, like in the proposed ViT for a fair comparison. The tested DeiT and Swin-v2 architectures are DeiT-S and Swin-v2-T, respectively (i.e., Small and Tiny), as being computationally efficient configurations, and also more suitable for the typical scenario of having a scarce number of samples. In addition, the Local InFormation Enhancer (LIFE) (\citet{akkaya2024enhancing}) module has been integrated into the DeiT-S and Swin-v2-T architectures to increase the receptive fields of their self-attention blocks by including patch-level local information. In this way, the proposed ViT with the joint SSL and supervised learning scheme is compared against deep neural network architectures enhanced with the LIFE module, as both introduce mechanisms to increase the locality inductive bias in ViTs.

\item Performance: Table \ref{tab:table_results_comp} summarizes the comparative performance in the heart failure time prediction task in terms of the error margin. This analysis includes the proposed system with either the ViT trained with the joint SSL and supervised learning scheme or with each alternative architecture: DeiT-S, Swin-v2-T, and ResNet50-v2 (both with and without pretraining). The most representative training cases from Table \ref{tab:training_cases} (specifically, the cases 1, 7, 10, 11, 12, 13, and 14) have been selected for these experiments. This choice rests on the favorable impact of incorporating SVD and HODMD-derived data instead of only the original echocardiography images for training. In the case of the test data, three types have been considered: original images, HODMD-based reconstructions and HODMD modes, which illustrate the contribution of the SVD and HODMD algorithms for feature extraction.

\begin{table*}[!ht]
\centering
\caption{Comparison of the heart failure time prediction performance in terms of the error margin among different deep neural networks using the most representative types of test data and training cases (best results in \textbf{bold}, in Case 14). Refer to Table~\ref{tab:training_cases} for the specifications of each training case.}
\label{tab:table_results_comp}
\resizebox{\textwidth}{!}{%
\begin{tabular}{@{}ccccccccc@{}}
\toprule
                                        &                       & \multicolumn{7}{c}{Error margin $\downarrow$}                                                                                                                                                                                                           \\ \midrule
Algorithm                               & Test Data             & Case 1                           & Case 7                           & Case 10                          & Case 11                          & Case 12                          & Case 13                          & Case 14                           \\ \midrule
\multirow{3}{*}{DeiT-S}                 & Original              & 4.81                             & 4.58                             & 4.64                             & 4.89                             & 5.09                             & 4.84                             & 5.43                              \\
                                        & HODMD Reconstructions & 4.87                             & 4.58                             & 4.66                             & 4.88                             & 5.19                             & 4.93                             & 5.46                              \\
                                        & HODMD Modes           & 4.73                             & 4.60                             & 4.64                             & 4.90                             & 4.85                             & 4.76                             & 5.28                              \\
\multirow{3}{*}{Swin-v2-T}                 & Original              & 5.07                             & 4.61                             & 4.62                             & 4.54                             & 4.79                             & 4.93                             & 4.54                              \\
                                        & HODMD Reconstructions & 5.09                             & 4.61                             & 4.62                             & 4.55                             & 4.83                             & 4.97                             & 4.54                              \\
                                        & HODMD Modes           & 7.91                                & 4.64                             & 4.60                             & 4.60                             & 4.60                             & 4.82                             & 4.62                              \\
\multirow{3}{*}{ResNet50-v2}            & Original              & 5.21                             & 4.77                             & 4.74                                & 4.78                             & 4.87                             & 4.86                             & 5.04                              \\
                                        & HODMD Reconstructions & 4.87                             & 4.77                             & 4.93                             & 4.95                             & 4.94                             & 4.94                             & 5.06                              \\
                                        & HODMD Modes           & 4.98                             & 4.86                             & 4.74                             & 4.74                             & 4.83                             & 5.01                             & 4.86                              \\
\multirow{3}{*}{ResNet50-v2-pretrained} & Original              & 5.20                             & 5.50                             & 4.81                             & 4.81                             & 4.56                             & 4.93                             & 5.01                              \\
                                        & HODMD Reconstructions & 4.87                             & 4.80                             & 4.71                             & 4.90                             & 4.91                             & 4.83                             & 4.96                              \\
                                        & HODMD Modes           & 7.06                                & 4.79                             & 4.73                             & 4.81                             & 4.77                             & 4.72                             & 4.83                              \\
\multirow{3}{*}{$\textbf{Proposed}$}    & Original              & 4.75                             & 4.57                             & 4.65                             & 4.59                             & 4.63                             & 4.56                             & $\mathbf{4.54}$                              \\
                                        & HODMD Reconstructions & 5.10                             & 4.57                             & 4.65                             & 4.60                             & 4.64                             & 4.56                             & 4.55                              \\
                                        & HODMD Modes           & 4.87                             & 4.84                             & 4.59                             & 4.75                             & 4.65                             & 4.71                             & $\mathbf{4.53}$                   \\ \cmidrule(l){2-9} 
                                        & \# Samples (train)    & \num[group-separator={~}]{27293} & \num[group-separator={~}]{38273} & \num[group-separator={~}]{49253} & \num[group-separator={~}]{66036} & \num[group-separator={~}]{77016} & \num[group-separator={~}]{93329} & \num[group-separator={~}]{104309} \\ \bottomrule
\end{tabular}%
}
\end{table*}

Following the results from the table, the lowest error margins are achieved with the proposed ViT and the Swin-v2-T. Both outperform ResNet50-v2, proving the benefits of using Transformers which incorporate mechanisms to enhance generalization under scarce-data scenarios. This is, this version of Swin-v2-T incorporates the LIFE module to increase the receptive field in the input of the attention mechanisms. In the case of the proposed ViT, this model leverages the joint learning scheme to improve representation learning even in the scenario of scarce samples. These results also prove the capability of these models to effectively capture complex non-linear relationship between echocardiography imaging and heart failure times. In general, the selected training cases improve the performance with respect to the case 1, and also the use of HODMD modes for test data instead of original images. Again, these results demonstrate the effectiveness of the SVD and especially HODMD algorithms to extract cardiac dynamics and the evolution of the heart state. Moreover, these results show the great potential of these Modal Decomposition techniques to filter the high noise inherent to the echocardiography imaging, and also other types of noise. Consequently, the tested alternative neural networks also benefit from the feature extraction and data augmentation performed by the SVD and HODMD algorithms. On the contrary, in the training cases without SVD nor HODMD modes (as in the case 1), the models tend to perform better on original test images than on the representations from the Modal Decomposition techniques. This behavior can be likely attributed to the fact that the models learn the features associated to the cardiac dynamics directly from the original echocardiography imaging, more accurately predicting when using this representation instead of the ones resulting from the Modal Decomposition techniques.

Overall, the proposed ViT trained with the MAE scheme and the joint SSL and supervised learning approach consistently has the best heart failure prediction performance. Among the tested alternative algorithms, the Swin-v2-T performs comparably. This proximity in the results can be attributed to the architectural and training features of the Swin-v2-T, which collectively mitigate the dependence on large databases for a proper training: the LIFE module (\citet{akkaya2024enhancing}), the mixup-based augmentation (which proved detrimental for the proposed ViT), and the application of shifted windows to enhance self-attention with local feature modeling. In any case, expanding the cardiac database could further benefit the proposed ViT and the alternative algorithms. Moreover, the results obtained with ResNet50-v2 highlight the limited advantage offered by transfer learning from natural image datasets like ImageNet for the heart disease recognition-based tasks like heart disease classification or heart failure prediction. This fact stems from the inherent differences between natural and medical imaging. In this case, the echocardiography imaging is standardized around specific views focusing on the region of the heart, and presents subtle fine textures and distinctive pixel intensity distributions different to those of natural images due to its acquisition conditions. Conversely, natural images emphasize global semantic structures and easily distinguishable shapes rarely relevant to the heart disease recognition-based tasks. As a result, pretrained models from natural image domains like those adopted in \citet{farhad2023data} usually fail to effectively address this kind of tasks.      

The results collected in Table \ref{tab:table_results_comp} also show that the best performances across the compared deep neural networks are obtained in the training cases incorporating HODMD-based data. For the proposed ViT, the training case 14 yields the lowest heart failure prediction error margin ($4.54$ months). This underscores the combined benefit of all types of data, validating the effectiveness of the cardiac database creation procedure proposed in this work and presented in Subsection \ref{sec:cardiac_data_creation}. This also demonstrates the capability to extend the methodology described in \citet{bell2025automatic} by generating a larger annotated database and by broadening the application range to the more complex and challenging task of heart failure time prediction. In general, the inclusion of original images or SVD-based data for training also contributes positively, though with less impact, as illustrated with the results obtained with the Swin-v2-T.

Table \ref{tab:comp_cost} presents a summary of the computational cost of the different phases involved in the prediction process. Note that the time values reported correspond to averages per image, although the SVD and the HODMD algorithms process sequences rather than individual snapshots. Also note that the these Modal Decomposition techniques, as data-driven methods, are independent of the rest of the framework, i.e, of the deep neural network and of the task to address. This explains the fact that the corresponding time results, i.e., $\bar{t}_{SVD}$, $\bar{t}_{HOSVD}$, and $\bar{t}_{HODMD}$, concord with those obtained for heart disease classification (\citet{bell2025automatic}). Among the steps involved in the SVD and the HODMD algorithms, the HOSVD-based dimensionality reduction is the most computationally demanding one (\citet{Bell2023HODMD}). Regarding the compared deep neural networks, the proposed ViT trained using the MAE scheme and the joint SSL and supervised learning strategy achieves one of the lowest prediction times and a much higher throughput than the one of Swin-v2-T. Compared with ResNet50-v2, the proposed ViT has a similar $\bar{t}_{pred}$, but achieves a superior trade-off between computational cost and prediction performance. In addition, the proposed ViT has the lowest number of learnable parameters among the compared architectures, supporting its suitability for deployment on devices with limited computational resources. However, all time values comply with real-time constraints, demonstrating the great potential of the proposed framework for several medical applications under different configurations, including the integration into portable diagnostic tools to cover remote areas.

Overall, the proposed ViT achieves the best trade-off between heart failure prediction performance and efficiency with a minimal computational overhead. This is accomplished using only the joint SSL and supervised learning scheme, without relying on external databases, unlike ResNet50-v2. This fact reduces training complexity and computational cost, and also mitigates the dependence on large high-quality annotated databases. However, further improvements in the heart failure prediction performance can be anticipated if larger echocardiography databases annotated with heart failure times were available.

\begin{table*}[!ht]
\centering
\caption{Computational cost of different phases of the Heart Failure Prediction System for the alternative algorithms.}
\label{tab:comp_cost}
\resizebox{\textwidth}{!}{
\begin{tabular}{@{}cccccccc@{}}
\toprule
Algorithm       & \# Parameters (M) & GFLOPs $\downarrow$ & $\bar{t}_{SVD} \downarrow$      & $\bar{t}_{HOSVD} \downarrow$    & $\bar{t}_{HODMD} \downarrow$      & $\bar{t}_{pred} \downarrow$ & Throughput $\uparrow$ \\ \midrule
DeiT-S      & 21.81         & 4.64   & \multirow{4}{*}{5.1} & \multirow{4}{*}{591} & \multirow{4}{*}{0.648} & 27.05            & 36.98      \\
Swin-v2-T   & 27.73         & 6.75   &                      &                      &                        & 18.96            & 52.84      \\
ResNet50-v2 & 23.51         & 4.11   &                      &                      &                        & 7.44             & 134.82     \\
\textbf{Proposed}    & 6.01          & 1.68   &                      &                      &                        & 9.71             & 105.76     \\ \bottomrule
\end{tabular}%
}
\end{table*}

\end{enumerate}

\section{Conclusions}
\label{sec:conclusions}

The leading cause of human defunction worldwide nowadays is heart diseases. In particular, heart failures have become predominant, which means that predicting specifically \emph{when} these happen results critical for early intervention and optimal treatment planning. However, this prediction remains a challenging problem mainly due to the scarcity of annotated echocardiography images and the complexity of the heart state evolution, and thus disease progression. This work presents a novel framework which extends our previous study from \citet{bell2025automatic}. This is because this paper proposes several changes to the methodology and broadens the application range from heart disease classification to the more challenging and demanding task of predicting the time in which a heart failure happens. This task has not been addressed in the related literature to the best of the authors' knowledge. Moreover, estimating the concrete age of a patient of a heart failure provides more accurate and relevant clinical information about that patient for personalized medicine applications than the more general and widely studied tasks in the related literature, namely the heart disease classification (like in \citet{bell2025automatic}), and the estimation of the grade of heart failure (low, high) in standard timelines. Among the extensions to our previous study, first, the database creation procedure proposed here generates a larger annotated database. This enables the combination of all types of data for training: the original echocardiography imaging, reconstructions, and modes. Therefore, the proposed creation process overcomes the limitations of our previous study, in which only a subset of the data could be leveraged. This creation process includes Modal Decomposition techniques, namely the SVD and HODMD algorithms, for both feature extraction and data augmentation, which are based on data physics, unlike conventional data augmentation techniques. Second, this work adopts MAEs, trained using a joint SSL and supervised learning scheme in a more efficient and effective way than the modified ViT from our previous work. To the best of the authors' knowledge, the general trend in the related literature on heart disease recognition and heart failure prediction in echocardiography images barely explores SSL, even less MAEs, which have precisely become the state of the art in Computer Vision. These two contributions, acting synergistically, successfully address the usual challenge of having a scarce number of samples in the medicine field, reducing the dependence on very large training datasets. More precisely, the elaboration of varied databases with high quality enough is very difficult, implying high costs and requiring very hard specialized work from experts. This fact leads to the necessity to adopt techniques specifically designed to address this problem, thus the aforementioned contributions. The results obtained have proved that the proposed system performs better than ResNet-v2 (even with pretraining), and also better than other established ViT-based approaches, which incorporate additional local information of patches with the LIFE module. In addition, the proposed deep neural network works in real-time and has a lower number of trainable parameters, so this is more efficient and well suited for several clinical applications, such as portable devices. We conclude that, if longer databases of echocardiography imaging or integrating different modalities (like clinical metadata) about heart failures were available, overfitting would be further reduced and robustness would be enhanced.

\section*{Declaration of competing interest}

The authors declare the following financial interests/personal relationships which may be considered as potential competing interests: Andrés Bell-Navas reports financial support was provided by Spanish Ministry of Science and Innovation. Enrique Lara-Pezzi reports financial support was provided by Spanish Ministry of Science and Innovation. Soledad Le Clainche reports financial support was provided by the Ministerio de Ciencia, Innovación y Universidades. Enrique Lara-Pezzi reports financial support was provided by Comunidad de Madrid. María Villalba-Orero reports financial support was provided by Comunidad de Madrid. María Villalba-Orero reports financial support was provided by Juan de la Cierva Incorporación. Enrique Lara-Pezzi reports a relationship with Spanish National Cardiovascular Research Center that includes: employment and funding grants. María Villalba-Orero reports a relationship with Spanish National Cardiovascular Research Center that includes: employment and funding grants. Given his role as Editor-in-Chief of Journal of Cardiovascular Translational Research, Enrique Lara-Pezzi had no involvement in the peer-review of this article and has no access to information regarding its peer-review. Full responsibility for the editorial process for this article was delegated to another journal editor. If there are other authors, they declare that they have no known competing financial interests or personal relationships that could have appeared to influence the work reported in this paper.

\section*{Acknowledgements}

The authors acknowledge the Grants TED2021-129774B-C21, TED2021-129774B-C22 and PLEC2022-009235, funded by MCIN/AEI/10.13039/\\501100011033 and by the European Union "NextGenerationEU"/PRTR, and the Grant PID2023-147790OB-I00, funded by MICIU/AEI/10.13039/\\501100011033/FEDER, UE: the first one to A.B-N, the next two to E.L-P, and the last one to S.L. The authors also acknowledge the Grant PEJ-2019-TL/BMD-12831 from Comunidad de Madrid to E.L-P and to M.V-O, and a Juan de la Cierva Incorporación Grant (IJCI-2016-27698) to M.V-O. The CNIC is supported by the Instituto de Salud Carlos III (ISCIII), the Ministerio de Ciencia, Innovación y Universidades (MICIU), and the Pro CNIC Foundation, and is a Severo Ochoa Center of Excellence (Grant CEX2020-001041-S funded by MICIU/AEI/10.13039/501100011033).

\section*{Data availability}

The data that have been used are confidential. For further details, please contact with Enrique Lara-Pezzi (elara@cnic.es).

\bibliographystyle{elsarticle-harv} 
\bibliography{bibliography}

@article{liu2023generalized,
  title={A generalized deep learning model for heart failure diagnosis using dynamic and static ultrasound},
  author={Liu, Zeye and Huang, Yuan and Li, Hang and Li, Wenchao and Zhang, Fengwen and Ouyang, Wenbin and Wang, Shouzheng and Luo, Zhiling and Wang, Jinduo and Chen, Yan and others},
  journal={J. Transl. Intern. Med.},
  volume={11},
  number={2},
  pages={138--144},
  doi={https://doi.org/10.2478/jtim-2023-0088},
  year={2023},
  publisher={De Gruyter}
}

@article{valsaraj2023development,
  title={Development and validation of echocardiography-based machine-learning models to predict mortality},
  author={Valsaraj, Akshay and Kalmady, Sunil Vasu and Sharma, Vaibhav and Frost, Matthew and Sun, Weijie and Sepehrvand, Nariman and Ong, Marcus and Equilbec, Cyril and Dyck, Jason RB and Anderson, Todd and others},
  journal={EBioMedicine},
  volume={90},
  doi={https://doi.org/10.1016/j.ebiom.2023.104479},
  year={2023},
  publisher={Elsevier}
}

@article{bell2025automatic,
  title={Automatic cardiac pathology recognition in echocardiography images using higher order dynamic mode decomposition and a vision transformer for small datasets},
  author={Bell-Navas, Andr{\'e}s and Groun, Nourelhouda and Villalba-Orero, Mar{\'\i}a and Lara-Pezzi, Enrique and Garicano-Mena, Jes{\'u}s and Le Clainche, Soledad},
  journal={Expert Syst. Appl.},
  volume={264},
  pages={125849},
  doi={https://doi.org/10.1016/j.eswa.2024.125849},
  year={2025},
  publisher={Elsevier}
}

@article{petmezas2024recent,
  title={Recent advancements and applications of deep learning in heart failure: A systematic review},
  author={Petmezas, Georgios and Papageorgiou, Vasileios E and Vassilikos, Vasileios and Pagourelias, Efstathios and Tsaklidis, George and Katsaggelos, Aggelos K and Maglaveras, Nicos},
  journal={Comput. Biol. Med.},
  pages={108557},
  doi={https://doi.org/10.1016/j.compbiomed.2024.108557},
  year={2024},
  publisher={Elsevier}
}

@article{zhang2018fully,
  title={Fully automated echocardiogram interpretation in clinical practice: feasibility and diagnostic accuracy},
  author={Zhang, Jeffrey and Gajjala, Sravani and Agrawal, Pulkit and Tison, Geoffrey H and Hallock, Laura A and Beussink-Nelson, Lauren and Lassen, Mats H and Fan, Eugene and Aras, Mandar A and Jordan, ChaRandle and others},
  journal={Circ.},
  volume={138},
  number={16},
  pages={1623--1635},
  doi={https://doi.org/10.1161/CIRCULATIONAHA.118.034338},
  year={2018},
  publisher={Lippincott Williams \& Wilkins Hagerstown, MD}
}

@article{akerman2023automated,
  title={Automated echocardiographic detection of heart failure with preserved ejection fraction using artificial intelligence},
  author={Akerman, Ashley P and Porumb, Mihaela and Scott, Christopher G and Beqiri, Arian and Chartsias, Agisilaos and Ryu, Alexander J and Hawkes, William and Huntley, Geoffrey D and Arystan, Ayana Z and Kane, Garvan C and others},
  journal={JACC Adv.},
  volume={2},
  number={6},
  pages={100452},
  doi={https://doi.org/10.1016/j.jacadv.2023.100452},
  year={2023},
  publisher={American College of Cardiology Foundation Washington DC}
}

@inproceedings{behnami2018automatic,
  title={Automatic detection of patients with a high risk of systolic cardiac failure in echocardiography},
  author={Behnami, Delaram and Luong, Christina and Vaseli, Hooman and Abdi, Amir and Girgis, Hany and Hawley, Dale and Rohling, Robert and Gin, Ken and Abolmaesumi, Purang and Tsang, Teresa},
  booktitle={Int. Workshop Deep Learn. Med. Image Anal.},
  pages={65--73},
  doi={https://doi.org/10.1007/978-3-030-00889-5_8},
  year={2018},
  organization={Springer}
}

@inproceedings{fazry2022hierarchical,
  title={Hierarchical vision transformers for cardiac ejection fraction estimation},
  author={Fazry, Lhuqita and Haryono, Asep and Nissa, Nuzulul Khairu and Hirzi, Naufal Muhammad and Rachmadi, Muhammad Febrian and Jatmiko, Wisnu and others},
  booktitle={2022 7th Int. Workshop Big Data Inf. Secur. (IWBIS)},
  pages={39--44},
  doi={https://doi.org/10.1109/IWBIS56557.2022.9924664},
  year={2022},
  organization={IEEE}
}

@inproceedings{muhtaseb2022echocotr,
  title={Echocotr: Estimation of the left ventricular ejection fraction from spatiotemporal echocardiography},
  author={Muhtaseb, Rand and Yaqub, Mohammad},
  booktitle={Med. Image Comput. Comput. Assist. Interv.}, 
  pages={370--379},
  doi={https://doi.org/10.1007/978-3-031-16440-8_36},
  year={2022},
  organization={Springer}
}

@inproceedings{reynaud2021ultrasound,
  title={Ultrasound video transformers for cardiac ejection fraction estimation},
  author={Reynaud, Hadrien and Vlontzos, Athanasios and Hou, Benjamin and Beqiri, Arian and Leeson, Paul and Kainz, Bernhard},
  booktitle={Med. Image Comput. Comput. Assist. Interv.--MICCAI 2021: 24th International Conference, Strasbourg, France, September 27--October 1, 2021, Proceedings, Part VI 24},
  pages={495--505},
  doi={https://doi.org/10.1007/978-3-030-87231-1_48},
  year={2021},
  organization={Springer}
}

@ARTICLE{Arooj2022deep,
  title={A Deep Convolutional Neural Network for the Early Detection of Heart Disease},
  author={Arooj, Sadia and Rehman, Saif ur and Imran, Azhar and Almuhaimeed, Abdullah and Alzahrani, A Khuzaim and Alzahrani, Abdulkareem},
  journal={Biomedicines},
  volume={10},
  number={11},
  pages={2796},
  doi={https://doi.org/10.3390/biomedicines10112796},
  year={2022},
  publisher={MDPI}
}

@ONLINE{who1999cvds,
  author = {{World Health Organization}},
  title = {Cardiovascular Diseases ({CVD}s)},
  year = {1999},
  url = {https://www.who.int/news-room/fact-sheets/detail/cardiovascular-diseases-(cvds)},
  note = {visited on 2021-06-11}
}

@ARTICLE{farhad2023data,
  title={A data-efficient zero-shot and few-shot {S}iamese approach for automated diagnosis of left ventricular hypertrophy},
  author={Farhad, Moomal and Masud, Mohammad Mehedy and Beg, Azam and Ahmad, Amir and Ahmed, Luai A and Memon, Sehar},
  journal={Comput. Biol. Med.},
  volume={163},
  pages={107129},
  doi={https://doi.org/10.1016/j.compbiomed.2023.107129},
  year={2023},
  publisher={Elsevier}
}

@ARTICLE{Lee2021vision,
  title={Vision transformer for small-size datasets},
  author={Lee, Seung Hoon and Lee, Seunghyun and Song, Byung Cheol},
  journal={arXiv preprint arXiv:2112.13492},
  year={2021}
}

@article{HETHERINGTON2024109217,
author = {Hetherington, Ashton and Corrochano, Adri{\'a}n and Abad{\'\i}a-Heredia, Rodrigo and Lazpita, Eneko and Mu{\~n}oz, Eva and D{\'\i}az, Paula and Maiora, Egoitz and L{\'o}pez-Mart{\'\i}n, Manuel and Le Clainche, Soledad},
title = {ModelFLOWs-app: Data-driven post-processing and reduced order modelling tools},
journal = {Comput. Phys. Commun.},
volume = {301},
pages = {109217},
year = {2024},
issn = {0010-4655},
doi = {https://doi.org/10.1016/j.cpc.2024.109217},
}

@article{le2017higher,
  title={Higher order dynamic mode decomposition},
  author={Le Clainche, Soledad and Vega, Jos{\'e} M},
  journal={SIAM J. Appl. Dyn.},
  volume={16},
  number={2},
  pages={882--925},
  doi={https://doi.org/10.1137/15M1054924},
  year={2017},
  publisher={SIAM}
}

@article{LECLAINCHE2017336,
title = {Higher order dynamic mode decomposition of noisy experimental data: The flow structure of a zero-net-mass-flux jet},
journal = {Exp. Therm. Fluid Sci.},
volume = {88},
pages = {336-353},
year = {2017},
issn = {0894-1777},
doi = {https://doi.org/10.1016/j.expthermflusci.2017.06.011},
url = {https://www.sciencedirect.com/science/article/pii/S089417771730184X},
author = {Soledad {Le Clainche} and José M. Vega and Julio Soria}
}

@article{schmid2010dynamic,
  title={Dynamic mode decomposition of numerical and experimental data},
  author={Schmid, Peter J},
  journal={J. Fluid Mech.},
  volume={656},
  pages={5--28},
  doi={https://doi.org/10.1017/S0022112010001217},
  year={2010},
  publisher={Cambridge University Press}
}

@article{sirovich1987turbulence,
  title={Turbulence and the dynamics of coherent structures. I. Coherent structures},
  author={Sirovich, Lawrence},
  journal={Q. Appl. Math.},
  volume={45},
  number={3},
  pages={561--571},
  year={1987}
}

@article{tucker1966some,
  title={Some mathematical notes on three-mode factor analysis},
  author={Tucker, Ledyard R},
  journal={Psychometrika},
  volume={31},
  number={3},
  pages={279--311},
  year={1966},
  publisher={Springer}
}

@ARTICLE{Groun2022higher,
  title={Higher order dynamic mode decomposition: From fluid dynamics to heart disease analysis},
  author={Groun, Nourelhouda and Villalba-Orero, Mar{\'\i}a and Lara-Pezzi, Enrique and Valero, Eusebio and Garicano-Mena, Jes{\'u}s and Le Clainche, Soledad},
  journal={Comput. Biol. Med.},
  volume={144},
  pages={105384},
  doi={https://doi.org/10.1016/j.compbiomed.2022.105384},
  year={2022},
  publisher={Elsevier}
}

@article{groun2025eigenhearts,
  title={Eigenhearts: Cardiac diseases classification using eigenfaces approach},
  author={Groun, Nourelhouda and Villalba-Orero, Mar{\'\i}a and Casado-Mart{\'\i}n, Luc{\'\i}a and Lara-Pezzi, Enrique and Valero, Eusebio and Le Clainche, Soledad and Garicano-Mena, Jes{\'u}s},
  journal={Comput. Biol. Med.},
  volume={192},
  pages={110167},
  doi={https://doi.org/10.1016/j.compbiomed.2025.110167},
  year={2025},
  publisher={Elsevier}
}

@INCOLLECTION{VEGA202129,
title = {Higher order dynamic mode decomposition},
editor = {Vega, Jos{\'e} M. and Le Clainche, Soledad},
booktitle = {Higher Order Dynamic Mode Decomposition and Its Applications},
publisher = {Academic Press},
pages = {29-83},
year = {2021},
isbn = {978-0-12-819743-1},
doi = {https://doi.org/10.1016/B978-0-12-819743-1.00009-4},
author = {Vega, Jos{\'e} M. and Le Clainche, Soledad}
}

@CONFERENCE{Bell2023HODMD,
  author  = {Bell-Navas, Andr{\'e}s and Groun, Nourelhouda and Garicano-Mena, Jes{\'u}s and Le Clainche, Soledad},
  title   = {Optimized Higher Order Dynamic Mode Decomposition Analysis of Electrocardiography Datasets},
  booktitle = {25th Conf. of ILAS}, 
  year    = {2023},
  pages   = {91-92}
}

@ARTICLE{ESAO2024,

  author  = {Bell-Navas, Andr{\'e}s and Groun, Nourelhouda and Villalba-Orero, Mar\'ia and Lara-Pezzi, Enrique and Garicano-Mena, Jes{\'u}s and Le Clainche, Soledad},
  title ={Automatic Heart Disease Prediction using Modal Decomposition and Masked Autoencoders for Limited Echocardiography Databases},
  journal = {Int. J. Artif. Organs},
  volume = {47},
  number = {7},
  pages = {471-472},
  doi = {https://doi.org/10.1177/03913988241279540},
  year = {2024}
}

@article{liu2025algorithms,
  title={Algorithms, Techniques and Applications of Intelligent Diagnosis Using Dynamic Ultrasound: A Review},
  author={Liu, Jialong and Zhang, Jianfeng and Shao, Yuan and Xiao, Xiaolong and Huang, Shoujun and Kong, Dexing},
  journal={IEEE Access},
  year={2025},
  doi={https://doi.org/10.1109/ACCESS.2025.3578817},
  publisher={IEEE}
}

@article{akkaya2024enhancing,
  title={Enhancing performance of vision transformers on small datasets through local inductive bias incorporation},
  author={Akkaya, Ibrahim Batuhan and Kathiresan, Senthilkumar S and Arani, Elahe and Zonooz, Bahram},
  journal={Pattern Recognit.},
  volume={153},
  pages={110510},
  doi = {https://doi.org/10.1016/j.patcog.2024.110510}, 
  year={2024},
  publisher={Elsevier}
}

@INPROCEEDINGS{he2016identity,
  title={Identity mappings in deep residual networks},
  author={He, Kaiming and Zhang, Xiangyu and Ren, Shaoqing and Sun, Jian},
  booktitle={Comput. Vis. ECCV 2016: 14th European Conference, Amsterdam, The Netherlands, October 11--14, 2016, Proceedings, Part IV 14},
  pages={630--645},
  doi={https://doi.org/10.1007/978-3-319-46493-0_38},
  year={2016},
  organization={Springer}
}

@inproceedings{touvron2021training,
  title={Training data-efficient image transformers \& distillation through attention},
  author={Touvron, Hugo and Cord, Matthieu and Douze, Matthijs and Massa, Francisco and Sablayrolles, Alexandre and J{\'e}gou, Herv{\'e}},
  booktitle={Int. Conf. Mach. Learn.},
  pages={10347--10357},
  year={2021},
  organization={PMLR}
}

@inproceedings{liu2022swin,
  title={Swin transformer v2: Scaling up capacity and resolution},
  author={Liu, Ze and Hu, Han and Lin, Yutong and Yao, Zhuliang and Xie, Zhenda and Wei, Yixuan and Ning, Jia and Cao, Yue and Zhang, Zheng and Dong, Li and others},
  booktitle={Proc. IEEE Comput. Soc. Conf. Comput. Vis. Pattern Recognit.},
  pages={12009--12019},
  doi={https://doi.org/10.1109/CVPR52688.2022.01170},
  year={2022}
}

@inproceedings{he2022masked,
  title={Masked autoencoders are scalable vision learners},
  author={He, Kaiming and Chen, Xinlei and Xie, Saining and Li, Yanghao and Doll{\'a}r, Piotr and Girshick, Ross},
  booktitle={Proc. IEEE Comput. Soc. Conf. Comput. Vis. Pattern Recognit.},
  pages={16000--16009},
  doi={https://doi.org/10.1109/CVPR52688.2022.01553},
  year={2022}
}

@inproceedings{das2024limited,
  title={Limited data, unlimited potential: A study on vits augmented by masked autoencoders},
  author={Das, Srijan and Jain, Tanmay and Reilly, Dominick and Balaji, Pranav and Karmakar, Soumyajit and Marjit, Shyam and Li, Xiang and Das, Abhijit and Ryoo, Michael S},
  booktitle={Proc. IEEE Winter Conf. Appl. Comput. Vis.},
  pages={6878--6888},
  doi={https://doi.org/10.1109/WACV57701.2024.00673},
  year={2024}
}

@ARTICLE{russakovsky2015imagenet,
  title={Image{N}et large scale visual recognition challenge},
  author={Russakovsky, Olga and Deng, Jia and Su, Hao and Krause, Jonathan and Satheesh, Sanjeev and Ma, Sean and Huang, Zhiheng and Karpathy, Andrej and Khosla, Aditya and Bernstein, Michael and others},
  journal={Int. J. Comput. Vis.},
  volume={115},
  pages={211--252},
  doi={https://doi.org/10.1007/s11263-015-0816-y},
  year={2015},
  publisher={Springer}
}





\end{document}